\documentclass[11pt,reqno]{article}

\usepackage{latexsym}
\usepackage{amsmath}
\usepackage{amsfonts,amssymb}

\usepackage{pict2e}

\usepackage{euscript}
\usepackage{eufrak}
\usepackage{cancel}
\usepackage{fancyhdr}

\begin{document}

\title{The  Ponzano-Regge asymptotic  \\ of the supersymmetric 6jS symbols}
\author{Lionel Br\'{e}hamet}



\maketitle
\begin{abstract}
\vspace{1em}
\noindent
We adapt the Gurau's proof (2008) about the asymptotic limit of Ponzano-Regge formula to supersymmetric 6jS symbols
according to their intrinsic parities alpha, beta, gamma.
The behaviour at a large scaling shows significant differences depending on these  parities. The decay is slowed 
and the angles of the oscillating parts in cosine are generally shifted or more altered.
Our results should be relevant in 3-D Quantum Supergravity and Spin Foam models. 
\\[1em]
\noindent 
PACS: 03.65.Fd  Algebraic methods\\  
PACS: 04.60.-m  Quantum gravity\\
PACS: 11.30.Pb  Supersymmetry  \\ [0.1em]
MSC: 81Q60; 83D05. \\[0.1em]
Keywords. 6j symbol; Quantum Gravity; supersymmetric 6jS  symbol.
\end{abstract}

\newpage
\section{Introduction}\label{Intro}\normalfont
\hspace*{1em}Since the growing expansion of quantum gravity theory \cite{Rovelli} the $SU(2)\,6j$ symbols acquired a considerable importance 
 by becoming the basic building blocks of all spin networks. They appear to represent a quantum tetrahedron with quantized edges and even can  be viewed as eigenfunctions of a discrete Schr\"{o}dinger equation \cite{Aquilanti}. As well known the classical Ponzano-Regge partition function $\mathcal{Z}_{PR}$ \cite{Ponzano}
was expressed as a sum of products of $6j$ symbols. In  \cite{Livine1} the occurrence of supersymmetric $6j^S$ symbols
 \cite{Daumensetal1,BrehametNuov1},
related to $OSP(1|2)$,  was also studied and led to a similar supersymmetric partition function, called  $\mathcal{Z}_{sugra}$.
A possible different divergence compared to the classical  case was sketched.\\
\hspace*{1em}Guided by the conceptual approach of Gurau \cite{Gurau1}, our present task is to give also {\sl an elementary proof} 
of the  Ponzano-Regge asymptotic of the $6j^S$ supersymmetric symbols. \\[0.2em]
Let us recall the formulation of Gurau:\\
\hspace*{1em}{\sl Under a rescaling of all its arguments by a large $k$ the $6j$ symbol associated to an Euclidean tetrahedron behaves like}
\begin{equation} \label{eq:General6jGurau}
\small
\left\{ \begin{array}{ccc} kj_1 &kj_2 & kj_{3} \\ kJ_1 & kJ_2 & kJ_3 \end{array} \right\}=\frac{1}{\sqrt{12\pi k^{3}V}}\cos
\Big\{ \frac{\pi}{4}+\sum_{\iota =1}^{3}\big[\big(kj_\iota+\frac{1}{2}\big)\theta_{j_{\iota}}+
\big(kJ_\iota+\frac{1}{2}\big)\theta_{J_{\iota}}\big] \Big\},
\end{equation}\normalsize
{\sl where $V$ is the tetrahedron volume and
 $\theta_{j_\iota}$, $\theta_{J_\iota}$ are the exterior dihedral angles of the tetrahedron corresponding to the edges $j_\iota$ and $J_\iota$ respectively.}\vspace{1em}

\setlength{\unitlength}{0.8cm}
\begin{center}
\begin{picture}(2.5,3.3)(-2.5,-0.5)
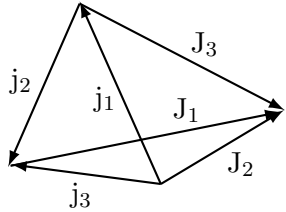
\put(-2.5,-1){{\scriptsize Tetrahedron of volume $V$}}
\thicklines
\put(-2.5,0.3){\vector(5,1){4.5}}
\put(0,0){\vector(5,3){2.05}}
\put(0,0){\vector(-23,3){2.5}}
\put(0,0){\vector(-4,9){1.35}}
\put(-1.35,3){\vector(-7,-16){1.19}}
\put(-1.35,3){\vector(25,-13){3.414}}
\put(1.1,0.2){\makebox{$\rm J_2$}}
\put(0.5,2.2){\makebox{$\rm J_3$}}
\put(0.2,1.1){\makebox{$\rm J _1$}}
\put(-1.1,1.3){\makebox{$\rm j_1$}}
\put(-2.5,1.6){\makebox{$\rm j_2$}}
\put(-1.5,-0.3){\makebox{$\rm j_3$}}
\end{picture}
\end{center}

				\section{Recalls about $6j$ and $6j^S$ symbols and notations}
We rewrite our formulas   \cite{BrehametNuov1} with notations as close as possible to those used by Gurau \cite{Gurau1} and
replace $j_i \leftrightarrow J_i, p_1\rightarrow v_4, p_2\rightarrow v_2, p_3\rightarrow v_3, p_4\rightarrow v_1,  q\rightarrow p $. Notations used in \cite{Gurau1}  were
\begin{align}
v_1=&j_1\!+\!j_2\!+\!j_3, \, v_2=J_1\!+\!j_2\!+\!J_3, \, v_3=J_1\!+\!J_2\!+\!j_3, \, v_4=j_1\!+\!J_2\!+\!J_3, \\
p_1=&j_2\!+\!J_2\!+\!j_3\!+\!J_3, \, p_2=j_3\!+\!J_3\!+\!j_1\!+\!J_1, \, p_3=j_1\!+\!J_1\!+\!j_2\!+\!J_2 .
\end{align}
\noindent
{\tt Diagrammatic representation of a $6j$ symbol}
$\left\{ \begin{array}{ccc} j_1 &j_2 & j_{3} \\ J_1 & J_2 & J_3 \end{array} \right\}$: \\
\hspace*{0.5em}
\vspace{1em}
\hspace*{0.1em}
\setlength{\unitlength}{0.4cm}
\linethickness{0.03pt}
\begin{picture}(27,3)
\put(7,2){\makebox(0,0){\scriptsize \bf  the four  triangles of any $6j$}}
\put(22,2){\makebox(0,0){\scriptsize \bf the three columns pairs}}
\put(0,0){\circle*{.135}} \put(1,1){\circle*{.135}} \put(2,1){\circle*{.135}}
\put(0,1){\circle{.35}} \put(1,0){\circle{.35}} \put(2,0){\circle{.35}}
\put(0,1){\line(1,-1){1}} \put(1,0){\line(1,0){1}}
\put(4,1){\circle*{.135}} \put(5,0){\circle*{.135}} \put(6,1){\circle*{.135}}
\put(4,0){\circle{.35}} \put(5,1){\circle{.35}} \put(6,0){\circle{.35}}
\put(4,0){\line(1,1){1}} \put(5,1){\line(1,-1){1}}
\put(8,1){\circle*{.135}} \put(9,1){\circle*{.135}} \put(10,0){\circle*{.135}}
\put(8,0){\circle{.35}} \put(9,0){\circle{.35}} \put(10,1){\circle{.35}}
\put(8,0){\line(1,0){1}} \put(9,0){\line(1,1){1}}
\put(12,0){\circle*{.135}} \put(13,0){\circle*{.135}} \put(14,0){\circle*{.135}}
\put(12,1){\circle{.35}} \put(13,1){\circle{.35}} \put(14,1){\circle{.35}}
\put(12,1){\line(1,0){1}} \put(13,1){\line(1,0){1}}
\put(17,0){\circle*{.135}} \put(18,0){\circle{.35}} \put(19,0){\circle{.35}}
\put(17,1){\circle*{.135}} \put(18,1){\circle{.35}} \put(19,1){\circle{.35}}
\put(18,1){\line(0,-1){1}} \put(19,1){\line(0,-1){1}}
\put(21,0){\circle{.35}} \put(22,0){\circle*{.135}} \put(23,0){\circle{.35}}
\put(21,1){\circle{.35}} \put(22,1){\circle*{.135}} \put(23,1){\circle{.35}}
\put(23,1){\line(0,-1){1}} \put(23,1){\line(0,-1){1}}
\put(25,0){\circle{.35}} \put(26,0){\circle{.35}} \put(27,0){\circle*{.135}}
\put(25,1){\circle{.35}} \put(26,1){\circle{.35}} \put(27,1){\circle*{.135}}
\put(25,1){\line(0,-1){1}} \put(26,1){\line(0,-1){1}}
\put(1,-1){\makebox(0,0){\scriptsize $v_4$}} \put(5,-1){\makebox(0,0){\scriptsize $v_2$}} \put(9,-1){\makebox(0,0){\scriptsize $v_3$}} \put(13,-1){\makebox(0,0){\scriptsize $v_1$}}
\put(18,-1){\makebox(0,0){\scriptsize $p_1$}} \put(22,-1){\makebox(0,0){\scriptsize $p_2$}} \put(26,-1)
{\makebox(0,0){\scriptsize $p_3$}}
\end{picture}				
\vspace{0.5em}\\
$v_i$ is the sum of the values of the three circled spins just above
$v_i$ in the diagrams. In the same way, $p_j$ is the sum of the values of  the four circled spins above $p_j$. The variables $p$,$v$ satisfy the equation
\begin{equation}\label {eq:IdBetweenpv}
\sum_{j=1}^3 p_j=\sum_{i=1}^4 v_i .
\end{equation}
As a result any spin is determined by two $v_i$ and one $p_j$ according to
\begin{align}
2j_1&=v_1+v_4-p_1, \,2j_2=v_1+v_2-p_2, \,2j_3=v_1+v_3-p_3,\\
2J_1&=v_2+v_3-p_1,\, 2J_2=v_3+v_4-p_2,\, 2J_3=v_2+v_4-p_3,
\end{align}
or by two $p_j$ and two $v_i$ `complementary' according to
\begin{align}
2j_1&=p_2 \!+\! p_3 \!-\!v_2 \!-\! v_3, \,2j_2=p_1 \!+\! p_3 \!-\! v_3 \!-\! v_4, \,2j_3=p_1 \!+\! p_2 \!-\! v_2 \!-\! v_4,\\
2J_1&=p_2 \!+\! p_3 \!-\! v_1 \!-\! v_4,\, 2J_2=p_1 \!+\! p_3 \!-\! v_1 \!-\! v_2,\, 2J_3=p_1 \!+\! p_2 \!-\! v_1 \!-\! v_3.
\end{align}
\noindent						
\hspace*{1em}Formulas for  a  $6j$ symbol are well known, however here we shall use an expression \cite{BrehametPJMPA}, used elsewhere long ago.
That avoids the usual triangle coefficients
 $\bigtriangleup$ and contains only terms in $v_i,p_j$.
\begin{equation} \label{eq:General6j1}
\left\{ \begin{array}{ccc} j_1 &j_2 & j_{3} \\ J_1 & J_2 & J_3 \end{array} \right\}= 
\sqrt{{\tt R}}
 \begin{array}{c}{\displaystyle \sum_{t}} \frac{(-1)^{t}(t+1)! } 
{\prod_{i=1}^{i=4}(t-v_i)! \prod_{j=1}^{j=3}( p_j-t)!  }\, ,\end{array}
\end{equation}
where the radical {\tt R} under square root is the prefactor used by Gurau
\begin{equation}\label{eq:PrefactorDef}
{\tt R}=\frac{\prod_{j=1}^{j=3}\prod_{i=1}^{i=4}(p_j-v_i)!}{\prod_{i=1}^{i=4}(v_i+1)!}.
\end{equation}
{\tt A short historical review about the supersymmetric $6j^{S}$ symbols:} \\The conceptualization of supersymmetric $3j$ or $6j$ symbols was done in several years spread over time,
with sometimes different notations. It started with Pais and Rittenberg \cite{PaisRittenberg} in 1975 with a semisimple graded Lie algebra named ``graded $\mathfrak{su(2)}$''. Later in 1977, Scheunert {\it et al.} \cite{ScheunertNahm} have computed the super-rotation Clebsch-Gordan coefficients as a product of usual rotation Clebsch-Gordan coefficients with a scalar factor. Finally, in 1981, Berezin and Tolstoy \cite{BerezinTostoy} suggested that the (iso)scalar factors form a pseudo-orthogonal matrix. Subsequently these works
were materialized in the paper by Daumens {\it et al.} \cite{Daumensetal1} dated 1993 with the introduction of properly defined  $6j^{S}$ symbols. \\ 
\hspace*{1em}Our definitions of supersymmetric $6j^{S}_{\pi}$ symbols \cite{BrehametNuov1} of parity $\pi=\alpha,\beta,\gamma$ were published in 2006, more than a decade after  the paper of Daumens {\it et al.}
\footnote{Ref.  \cite{Daumensetal1} was  explicitly used in 2004 by Livine and Oeckl \cite{Livine1} . On that date our work \cite{BrehametNuov1} was not known. }. 
\\ These  $6j^{S}_{\pi}$ symbols were expressed by a single sum formula over an index $t$ [involving monomials $ \Pi_{\pi}(t)$] as shown below:
\begin{equation}\label{GeneralFormula}
\begin{array}{c}
\left\{ \begin{array}{ccc} j_1 &j_2 & j_{3} \\ J_1 & J_2 & J_3 \end{array} \right\}_{\pi}^{S}=(-1)^{4\sum j_\iota J_\iota}
\sqrt{{\tt R}^{S}_{\pi}}\end{array}
\begin{array}{c} {\displaystyle  \! \sum_{t}} \frac{(-1)^{t}t! \Pi_{\pi}(t) } 
{\prod_{i=1}^{i=4}\left( t-[v_i+\frac{1}{2}]\right)! \prod_{j=1}^{j=3}
\left( [p_j+\frac{1}{2}]-t)\right)! }\, ,\end{array}
\end{equation}
where ${\tt R}^{S}_{\pi}$ is a supersymmetric prefactor given by 
\begin{equation}
{\tt R}^{S}_{\pi}=\frac{\prod_{j=1}^{j=3}\prod_{i=1}^{i=4}\boldsymbol{[}p_j-v_i\boldsymbol{]}!}{\prod_{i=1}^{i=4}\boldsymbol{[}v_i+\frac{1}{2}\boldsymbol{]}!}.
\end{equation}
Detailed expressions for ${\tt R}^{S}_{\alpha}$, ${\tt R}^{S}_{\gamma}$, ${\tt R}^{S}_{\beta}$ are given in Appendix.
Notation $\boldsymbol{[}\;\boldsymbol{]}$ means 'integer part of'.\\
In contrast to $SU(2)$,  $OSP(1|2)$ triangles $v_i$ can be integer or half-integer, however they still satisfy the  well known triangular inequalities.\\
Let us recall the definitions of the parities $\pi$ and monomials  $\Pi_{\pi}(t)$ of degree $0$ or $1$ in $t$:
\begin{equation}   \label{eq:parities}
\begin{cases}
\pi=\alpha   \text{ if }   \forall i\in [1,4] \text{\hspace{0.2em}} v_i  \text{\hspace{0.2em}integer},\\
\pi=\beta    \text{ if} \;\exists \, \text{\,only two distinct }v_i, v_j \text{\hspace{0.2em}integer},\\
\pi=\gamma   \text{ if}  \; \forall i\in [1,4] \text{\hspace{0.2em}} v_i \text{\hspace{0.2em}half-integer}.
\end{cases}
\end{equation}
For a  parity $\boldsymbol{\beta}$, both integer triangles shall be denoted by $v,v'$,
both other half-integer by  $\overline{v}, \overline{v}'$. The single integer quadrangle is denoted by $p$, 
both other half-integer  by $\overline{p}, \overline{p}'$.\\ 
In this case eq. (\ref{eq:IdBetweenpv}) transforms into 
\begin{equation}\label {eq:IdBetweenpvBeta}
p+\overline{p}+\overline{p}'=v+v'+\overline{v}+\overline{v}'.
\end{equation}
Thus a  parity $\pi$ just depends on the quality half-integral or integer of a triangle $v_i$. Let us note that only the $\gamma$ parity may contain
supersymmetric  symbols whose all the spins are half-integers.\\
Our monomials $\Pi_{\pi}(t)$ were defined in \cite{BrehametNuov1} as   
\begin{align}
 \Pi_{\alpha}(t)=& 1,   \label{eq:PiMonomS1}\\
\Pi_{\beta}(t)=& \begin{array}{c}-t(2 \mathfrak{j} ^{\star}_{\beta}+1)+(\overline{p}+\frac{1}{2})(\overline{p}' +\frac{1}{2})- vv' \end{array},   \label{eq:PiPiMonomS2}\\
\Pi_{\gamma}(t)=&-t\!+\!2(J_1j_1\!+\!J_2j_2\!+\!J_3j_3)\!+\!(J_1\!+\!j_1\!+\!J_2\!+\!j_2\!+\!J_3\!+\!j_3)\!+\!\frac{1}{2}. \label{eq:PiPiMonomS3}
\end{align}
All constants $\Pi_{\alpha }(0)$, $\Pi_{\beta}(0) $, $\Pi_{\gamma}(0) $ are positive {\tt integers}.\\
The special spin $\mathfrak{j} ^{\star}_{\beta}$ was identified \cite {BrehametNuov1} as the {\sl vertex} common to both half-integer triangles 
$\overline{v}, \overline{v}'$:
\begin{equation}
2 \mathfrak{j} ^{\star}_{\beta}= \overline{p}+\overline{p}' -v-v' = \overline{v}+\overline{v}' -p.
\end{equation}
Other shorter formulas are available: 
\begin{equation}
J_1j_1+J_2j_2+J_3j_3=\sum_{i=\iota}^{3} j_\iota J_\iota \quad \text{and} \quad J_1+j_1+J_2+j_2+J_3+j_3 = \frac{1}{2}\sum_{j=1}^{3}p_j .
\end{equation}
Oddly enough these `supersymmetric' quantities are reflected in most parameters of the discriminant necessary for the saddle points computation \cite{Gurau1},
while the background is that of standard $6j$.\\[0.2em]
\noindent
An appropriate  formulation more adapted to the manipulations of $\sum_t$ in \cite{Gurau1} is the following
\begin{align}
\Pi_{\beta}(t)=& \begin{array}{c}-(t+1)(2 \mathfrak{j} ^{\star}_{\beta}+1)
+[(2 \mathfrak{j} ^{\star}_{\beta}+1)+(\overline{p}+\frac{1}{2})
(\overline{p}' +\frac{1}{2})- vv' ]\end{array},   \label{PiMonomS2}\\
\Pi_{\gamma}(t)=&-(t+1)+[2\sum_{\iota=1}^{3} j_\iota J_\iota+ \frac{1}{2}\sum_{j=1}^{3}p_j +\frac{3}{2}]\label{PiGamma} . 
\end{align}
The supersymmetric frontal phase $(-1)^{4\sum j_\iota J_\iota}$ is worthwhile of attention. From  eq. (21) in \cite{Gurau1}
\begin{equation}
A = 2\sum_{\iota=1}^{3} j_\iota J_\iota=\sum_{i<j}v_iv_j -\sum_{k<l}p_k p_l ,
\end{equation}
it can be proved hat 
\begin{equation}
4\sum j_\iota J_\iota =
\begin{cases}
0\,  \pmod{2}\;\text{if}\;\pi=\alpha ,\\
1+\sum_j p_j  \pmod{2}\;\text{if}\;\pi=\gamma ,\\
v+v'-p \pmod{2}\;\text{if}\;\pi=\beta .
\end{cases}
\end{equation}
As $(-1^{k^{2}N}=+1$ if $k$ even and $(-1)^{N}$ if $k$ odd, rescaling any spin by  $\tt k$ {\tt odd} leads to
\begin{equation}
{(-1)^{4k^{2}\sum j_\iota J_\iota }}_{|\text{\tt k odd}}=
\begin{cases}
+1\;\text{if}\;\pi=\alpha ,\\
(-1)^{(1+\sum_j p_j )}\;\text{if}\;\pi=\gamma ,\\
(-1)^{(v+v'-p)}\;\text{if}\;\pi=\beta .
\end{cases}
\end{equation}
In the same way if  $v_i$ is integer,  $kv_i$ remains integer, whereas if $v_i$ is half-integer $kv_i$ remains
 half-integer only if  $k$ is odd and the formulas for initial parities $\beta,\gamma$ can change.
Thus care should be taken of $k$ parity itself.
Indeed the following transformations can occur: 
\begin{equation}
\left\{ \begin{array}{ccc} j_1 &j_2 & j_{3} \\ J_1 & J_2 & J_3 \end{array} \right\}_{\boldsymbol{\alpha}}^{S}
\stackrel{\times k}\longrightarrow 
\left\{ \begin{array}{ccc} kj_1 &kj_2 & kj_{3} \\ kJ_1 & kJ_2 & kJ_3 \end{array} \right\}_{\boldsymbol{\alpha}}^{S}
\quad \forall k \;\text{even or} \;\text{odd},\\
\end{equation} 
\begin{equation}
\left\{ \begin{array}{ccc} j_1 &j_2 & j_{3} \\ J_1 & J_2 & J_3 \end{array} \right\}_{\boldsymbol{\beta},\boldsymbol{\gamma}}^{S}
\stackrel{\times k}\longrightarrow 
\left\{ \begin{array}{ccc} kj_1 &kj_2 & kj_{3} \\ kJ_1 & kJ_2 & kJ_3 \end{array} \right\}_{\boldsymbol{\alpha}}^{S}
\quad \text{if}\; k\;\text{even},\\
\end{equation} 
\begin{equation}
\left\{ \begin{array}{ccc} j_1 &j_2 & j_{3} \\ J_1 & J_2 & J_3 \end{array} \right\}_{\boldsymbol{\beta},\boldsymbol{\gamma}}^{S}
\stackrel{\times k}\longrightarrow 
\left\{ \begin{array}{ccc} kj_1 &kj_2 & kj_{3} \\ kJ_1 & kJ_2 & kJ_3 \end{array} \right\}_{\boldsymbol{\beta},\boldsymbol{\gamma}}^{S}
\quad \text{if}\; k\;\text{odd}.
\end{equation}
The precisions given about `subleading contributions' in  \cite{Gurau1}	are not altered by the fact that $k \rightarrow \infty$ by even or odd integer values.\\
Due to the presence of integer parts, in ${\tt R}^{S}_{\pi}$ and   $\sum_t$, the Stirling's approximation 
\begin{equation}\label{eq:Stirling}
n!= \sqrt{2\pi }e^{(n+\frac{1}{2})\ln(n) - n}= \sqrt{2\pi }e^{n\ln(n) - n+\frac{1}{2}\ln(n)},
\end{equation}
could  give rise to different formulas when all spins are rescaled by a factor $k$.  Namely
\begin{equation}
\boldsymbol{[}kn \boldsymbol{]}! = (kn)!=\sqrt{2\pi}\;e^{[k\ln(k)-k]n +kn\ln(n)+\frac{1}{2}\ln(kn)} \quad  \text{if }n \; \text{integer}.\label{Approx1}
\end{equation}
\begin{equation}   \label{Approx2}
\boldsymbol{[}kn \boldsymbol{]}! = 
\begin{cases}
(kn)! \qquad \text{ if }k \; \text {even and }  n \text{ half-integer},\\
(kn-\frac{1}{2})! \,\text{ if }k \; \text {odd and }  n \text{ half-integer}.
\end{cases}
\end{equation}
\begin{equation}   \label{Approx3}
\boldsymbol{[}kn+\frac{1}{2} \boldsymbol{]}! = 
\begin{cases}
(kn)! \qquad \text{ if }k \; \text {even and }  n \text{ half-integer},\\
(kn+\frac{1}{2})! \,\text{ if }k \; \text {odd and }  n \text{ half-integer}.
\end{cases}
\end{equation}
It can be checked that only two approximations are modified for $k$ odd and $n$ half-integer:
\begin{align}
\boldsymbol{[}kn \boldsymbol{]}! =& \sqrt{2\pi}\;e^{[k\ln(k)-k]n +kn\ln(n)}, \label{Approx2_1}\\
\boldsymbol{[}kn +\frac{1}{2} \boldsymbol{]}! =& \sqrt{2\pi}\;e^{[k\ln(k)-k]n +kn\ln(n)+\ln(kn)}\label{Approx3_1}.
\end{align}
As a direct consequence of eq. (\ref{eq:IdBetweenpv}), we have single out the term $ e^{[k\ln(k)-k]}$ because in all the summations  to carry out with parities $\alpha, \beta, \gamma$ the resulting multiplicative factor will yield  a null quantity.\\
Since we look for a dominant behaviour  in first order approximation, the equalities are valid up to a multiplicative factor $1+1/k$. 
 Throughout this paper we will keep the original  functions of Gurau $H,h,F,f$, denoted here  by Gothic letters as  $\mathfrak{H},\mathfrak{h},\mathfrak{F},\mathfrak{f}$.\\
{\tt A  preliminary analysis shows that, for the three supersymmetric \\ parities $\alpha, \beta, \gamma$,  
both  functions $\mathfrak{h}(j,J)$ and $\mathfrak{f}(x)$ remain unchanged as \\well  as the saddle points values $x_{\pm}$}. \\
These functions appear under an exponentiated form as $e^{k\mathfrak{h}(j,J)}$ and $e^{k\mathfrak{f}(x)}$ . The results were
presented as a linear expansion over the six spins as
\begin{align} \label{eq1}
\mathfrak{h}(j,J) =& j_1\mathfrak{h}_{j_1}+j_2\mathfrak{h}_{j_2}+j_3\mathfrak{h}_{j_3}+ J_1\mathfrak{h}_{J_1}
+J_2\mathfrak{h}_{J_2}+J_3\mathfrak{h}_{J_3}, \\
\mathfrak{h}_{j_1}  =& \begin {array}{l} \frac{1}{2}\ln
\left\{ \frac{(j_1\!+\!j_2\!-\!j_3)(j_1\!-\!j_2\!+\!j_3)(j_1\!+\!J_2\!-\!J_3)(j_1\!-\!J_2\!+\!J_3)}
{(j_1+j_2+j_3)(-j_1+j_2+j_3)(j_1+J_2+J_3)(-j_1+J_2+J_3)}\right\}\end{array}\;\text{and so on},\label{hequation}\\
 \mathfrak{f}(x)=&\imath\pi x\!+\!x\ln(x)\!-\!\sum(x\!-\!v_i)\ln(x\!-\!v_i)-\sum(p_j\!-\!x)\ln(p_j\!-\!x). 
\end{align}
Equation (\ref{hequation}) can be rewritten in terms of $p,v$: 
\begin{align}
\mathfrak{h}_{j_1}  =&
\frac{1}{2}\ln \left\{\frac{(p_3-v_3)(p_2-v_2)(p_3-v_2)(p_2-v_3)}{(p_1-v_4)(p_1-v_1)v_1v_4}\right\}.
\end{align}
So we have
\begin{align}
h_\alpha(j,J) =& \, h_\beta(j,J) =h_\gamma(j,J) \equiv \mathfrak{h }(j,J),\\
h^\alpha_{j_1}=&\,h^\beta_{j_1}=h^\gamma_{j_1}\equiv \mathfrak{h}_{j_1}\;\text{and so on}.\\
f_{\alpha}(x)=&\,f_{\beta}(x)=f_{\gamma}(x)\equiv \mathfrak{f}(x).
\end{align}
Equation (18) in \cite{Gurau1} is the saddle points equation and remains unchanged, then the same holds for the saddle points solutions $x_{\pm}$ 
lying in the appropriate integration interval. We recall that
\begin{equation}
\mathfrak{f}^{'}(x)=\imath \pi \ln(x)-\sum \ln(x-v_i) +\sum \ln(p_j-x)=0.
\end{equation}
Changes with respect to the usual $6j$ symbols can arise either  from  ${\tt R}^{S}_{\pi}$
 {\it via}  $e^{H_{\pi}((j,J)\neq \mathfrak{H}((j,J)}$ or from   $\Sigma_x$ {\it via}  $e^{F_{\pi}(x)\neq \mathfrak{F}(x)}$ 
. This will lead  to distinct variants of Gurau's proof.

	\section {{\tt PARITY} $\boldsymbol{\alpha}$}\label{SectAlpha}
\renewcommand{\theequation}{\ref {SectAlpha}.\arabic{equation}}
\setcounter{equation}{0}
\subsection{Change in the prefactor ${\tt R}^{S}_{\alpha}$}
Multiplying the spins of the prefactor ${\tt R}^{S}_{\alpha}$ by a scaling factor $k$ yields
\begin{equation}
{\tt R}^{S}_{\alpha}(k)=\frac{\prod_{j=1}^{j=3}\prod_{i=1}^{i=4}\big(k(p_j-v_i)\big)!}{\prod_{i=1}^{i=4}(kv_i)!},
\end{equation}
From eq. (\ref{Approx1}) the numerator and denominator of ${\tt R}^{S}_{\alpha}(k)$ become
\begin{align}
{\tt NumPre\alpha}=&(2\pi)^{6}
e^{[k\ln(k)-k](4\sum_j p_j-3\sum_i v_i)+k[\sum_{j,i}(p_j-v_i)\ln(p_j-v_i)] }\nonumber \\
&\times e^{\frac{1}{2}\sum_{j,i}\ln(p_j-v_i)+\frac{1}{2}\sum_{j,i}\ln(k)}\nonumber \\
=&(2\pi)^{6}
e^{[k\ln(k)-k](4\sum_j p_j-3\sum_i v_i)+k[\sum_{j,i}(p_j-v_i)\ln(p_j-v_i)] }\nonumber \\
&\times e^{\frac{1}{2}\sum_{j,i}\ln(p_j-v_i)+6\ln(k))}.
\end{align}
\begin{align}
{\tt DenPre\alpha} =&(2\pi)^{2}
e^{[k\ln(k)-k]\sum_i v_i +k\sum_i v_i\ln(v_i)+\frac{1}{2}\sum_i \ln(k)+ \frac{1}{2}\sum_i \ln(v_i)}\nonumber \\
=&(2\pi)^{2}
e^{[k\ln(k)-k]\sum_i v_i +k\sum_i v_i\ln(v_i))}\nonumber \\
&\times e^{ \frac{3}{2}\sum_i \ln(v_i)-\sum_i \ln(v_i)+2\ln(k)} \nonumber \\
=&\frac{(2\pi)^{2}}{\Pi_i (v_i)}\,e^{[k\ln(k)-k]\sum_i v_i +k\sum_i v_i\ln(v_i)+\frac{1}{2}\sum_i \ln(v_i^{3})+2\ln(k)}.
\end{align}
That is
\begin{align}
{\tt R}^{S}_{\alpha}(k)=&(2\pi)^{4}k^{4}\Pi_i (v_i)\,e^{k[\sum_{j,i}(p_j-v_i)\ln(p_j-v_i)-\sum_i v_i\ln(v_i)}]\nonumber \\
&\times e^{\frac{1}{2}\sum_{j,i}\ln \frac{(p_j-v_i)}{v_i^{3}}}.
\end{align}
The similarities with eqs. (8-9) in \cite{Gurau1} are obvious whence 
\begin{equation}\label{eq:PrefactAlphaFinal}
\sqrt{{\tt R}^{S}_{\alpha}(k)}=
(2\pi)^{2}k^{2}\sqrt{\Pi_i (v_i)}\,e^{\mathfrak{H}(j,J)+k\mathfrak{h}(j,J)}
\end{equation}

			\subsection {Changes in the factors $\Sigma_{\boldsymbol{\alpha}}$ and $F_{\alpha}(x)$}\label{SFactorSigmaAlpha}
\begin{equation}
\Sigma_{\boldsymbol{\alpha}}=
\sum_{\max v_i}^{\min p_j}\frac{(-1)^{t}t!}{\Pi_{i}(t-v_i)!\Pi_{j}(p_j-t)!}.
\end{equation}
The only difference with \cite{Gurau1} lies in the lack of factor $e^{\ln(kx+1))}\approx e^{\frac{1}{2}\ln(k^{2}x^{2})}$ whence
\begin{equation}
\Sigma_{\boldsymbol{\alpha}}=\frac{1}{(2\pi)^{3}}\sum_{x=\max v_i}^{x=\min p_j}G_{\alpha}(x),
\end{equation}
with
\begin{align}
G_{\alpha}(x)=&
\frac{1}{k^{3}}e^{\frac{1}{2}\ln \frac{x}{\Pi(x-v_i)\Pi(p_j-x)}}\nonumber \\
&\times e^{k\{\imath \pi x+ x\ln(x)-[\sum_i(x-v_i)\ln(x-v_i)+\sum_j(p_j-x)\ln(p_j-x)]\}},
\end{align}
so that
\begin{equation}
\Sigma_{\boldsymbol{\alpha}}=\frac{1}{(2\pi)^{3}}\sum_{x=\max v_i}^{x=\min p_j}\frac{1}{k^{3}}
e^{F_{\alpha}(x)+k\mathfrak{f}(x)}.
\end{equation}
with
\begin{equation}
F_{\alpha}(x)=\frac{1}{2}\ln \frac{x}{\Pi(x-v_i)\Pi(p_j-x)}=\mathfrak{F}(x)-\ln(x).
\end{equation}
Accordingly
\begin{equation}\label{sumAlpha}
\Sigma_{\boldsymbol{\alpha}}=\frac{1}{(2\pi)^{3}}\frac{1}{k^{3}}\sum_{x=\max v_i}^{x=\min p_j}
\frac{1}{x}\,e^{\mathfrak{F}(x))+k\mathfrak{f}(x)}.
\end{equation}
			\subsection {Contribution to the saddle points (parity $\alpha$)}\label{ContribSaddleAlpha}

Identifying $\Sigma_{\boldsymbol{\alpha}}$ of eq. (\ref{sumAlpha}) as a Riemann sum with $\frac{1}{k} \rightarrow dx $ leads to
\begin{equation}\label{eq:SaddleContribSigmaAlpha}
\Sigma_{\boldsymbol{\alpha}}=\frac{1}{(2\pi)^{3}k^{2}}\int_{\max v_i}^{\min p_j}dx \frac{1}{x}\,e^{\mathfrak{F}(x)+k\mathfrak{f}(x)}.
\end{equation}
After collecting the three header factors  from eqs. (\ref{eq:PrefactAlphaFinal}, \ref{eq:SaddleContribSigmaAlpha}) and from the Laplace's approximation [in complex plane] we find
\begin{equation}
 {\tt Sdl}_{\alpha}(x_s)=
\displaystyle \frac{\sqrt{\Pi v_i}}{\sqrt{2\pi k}}\frac{1}{x_s}\frac{1}{\sqrt{-f''(x_s)}}
\, e^{\mathfrak{H}(j,J)+\mathfrak{F}(x_s)+k[\mathfrak{h}(j,J)+\mathfrak{f}(x_s)]},
\end{equation}
with the same values $x_s$ than those used by Gurau. For reducing the equation lengths let  us define
\begin{equation}\label{eq:PHIDef}
\varphi_{k}^{\theta}=\sum_{\iota=1}^{\iota=3}(kj_\iota+\frac{1}{2})\theta_{j_\iota}+(kJ_\iota+\frac{1}{2})\theta_{J_\iota}.
\end{equation}
Then eq. (45) in \cite{Gurau1} changes into
\begin{align}
&\frac{\displaystyle \sqrt{\Pi_{i=1}^{i=4} (v_i})}{\sqrt{48\pi kV}}\Big[\frac{1}{x_{+}}
e^{\imath\frac{\pi}{4}+\imath \varphi_{k}^{\theta}} 
+\frac{1}{x_{-}}
e^{-\imath\frac{\pi}{4}-\imath \varphi_{k}^{\theta}} 
\Big] \nonumber \\
&=\frac{\sqrt{\Pi v_i}}{\sqrt{48\pi kV}}
\Big[\big[ \frac{1}{x_{+}}+\frac{1}{x_{-}}\big]\cos\big(\frac{\pi}{4}+\varphi_{k}^{\theta}\big)
+\imath\big[ \frac{1}{x_{+}}-\frac{1}{x_{-}}\big]\sin\big(\frac{\pi}{4}+\varphi_{k}^{\theta}\big)\Big] .
\end{align}
By using the results and notations of  \cite{Gurau1} 
we  have
\begin{align}
&\frac{1}{x_{\pm}}= 
\frac{2A}{(B\pm \imath \sqrt{\bigtriangleup)}}\nonumber \\
&=\frac{2A(B\mp \imath \sqrt{\bigtriangleup)}}{B^2+\bigtriangleup}=
\frac{2A(B\mp \imath \sqrt{\bigtriangleup)}}{4AC}= \frac{(B\mp \imath \sqrt{\bigtriangleup)}}{2C},
\end{align}
whence
\begin{equation}
\frac{1}{x_{+}}+\frac{1}{x_{-}}=\frac {B}{C}\quad \text{and}\quad \frac{1}{x_{+}}-\frac{1}{x_{-}}=
 -\frac{\imath \sqrt{\bigtriangleup}}{C}.
\end{equation}
As $C=\Pi  v_i$ and $\sqrt{\bigtriangleup}=24V$ we obtain finally
\begin{equation}\label{eq:6jSAlphaApprox}
\frac{1}{\sqrt{48\pi kV}\sqrt{\Pi v_i}}
\Big[ B \cos\left(\frac{\pi}{4}+\varphi_{k}^{\theta}\right)
+\,24V\sin\left(\frac{\pi}{4}+\varphi_{k}^{\theta}\right)\Big].
\end{equation}
We recall that coefficient  $B$ defined in  \cite{Gurau1} has the dimension of a volume and writes as
\begin{equation} \label{eq:BRedef}
B=\Big(\sum_{i=1}^{3}j_i J_i \sum_{j=1}^{3}p_j \Big)+2(j_1j_2j_3 +j_1J_2J_3 +j_2J_3J_1 +j_3J_1J_2).
\end{equation}
The square root $\sqrt{\Pi v_i}$ has the dimension of an area.
		\section {{\tt PARITY} $\boldsymbol{\gamma}$ and $k$ odd}\label{SectGamma}
\renewcommand{\theequation}{\ref {SectGamma}.\arabic{equation}}
\setcounter{equation}{0}
			\subsection{Change in the prefactor ${\tt R}^{S}_{\gamma}$}
Multiplying the spins of the prefactor ${\tt R}^{S}_{\gamma}$ by a scaling odd factor $k$ yields
\begin{equation}
{\tt R}^{S}_{\gamma}(k)=\frac{\prod_{j=1}^{j=3}\prod_{i=1}^{i=4}\big(k(p_j-v_i)-\frac{1}{2})!}{\prod_{i=1}^{i=4}(kv_i+\frac{1}{2})!},
\end{equation}
\begin{equation}
{\tt NumPre\gamma}=(2\pi)^{6}
e^{[k\ln(k)-k](4\sum_j p_j-3\sum_i v_i)+k[\sum_{j,i}(p_j-v_i)\ln(p_j-v_i)]}.
\end{equation}
\begin{align}
{\tt DenPre\gamma}&=(2\pi)^{2}
e^{[k\ln(k)-k]\sum_i v_i+\sum_{i}\ln(k)+\sum_{i}\ln(v_i)}\nonumber \\
&=(2\pi)^{2}k^{4}e^{[k\ln(k)-k]\sum_i v_i+\sum_{i}\ln(v_i)}.
\end{align}
Then
\begin{equation}
\sqrt{{\tt R}^{S}_{\gamma}(k)}=(2\pi)^{2}\frac{1}{k^{2}}
e^{\frac{1}{2}k[\sum_{j,i}(p_j-v_i)\ln(p_j-v_i)-\sum_{i}\ln(v_i)]}.
\end{equation}
This means that the prefactor $\sqrt{{\tt R}^{S}_{\gamma}(k)}$ is given by
\begin{equation}\label{eq:FrontalFactGamma'1}
\sqrt{{\tt R}^{S}_{\gamma}(k)}=(2\pi)^{2}\frac{1}{k^{2}}
e^{H_{\gamma}(i,J)+kh_{\gamma}(i,J)},
\end{equation}
with
\begin{equation}
H_{\gamma}(i,J)=0 \quad \text{and}\quad h_{\gamma}(i,J)\equiv \mathfrak{h}(i,J).
\end{equation}

			\subsection {Changes in the factors $\Sigma_{\boldsymbol{\gamma}}$ and $F_{\gamma}(x)$}\label{SFactorSigmaGamma}
From eqs. (\ref{GeneralFormula})  and (\ref{PiGamma}) let be two sums $\Sigma^{'}_{\boldsymbol{\gamma }}
,\Sigma^{''}_{\boldsymbol{\gamma }}$
to be gathered to form a $\Sigma_{\boldsymbol{\gamma}}$:
\begin{align}
\Sigma^{'}_{\boldsymbol{\gamma }}= &
\sum_{t=\max v_i+\frac{1}{2}}^{t=\min p_j}\frac{(-1)^{t}(t+1)!}{\Pi_{i}(t-v_i\!-\!\frac{1}{2})!\Pi_{j}(p_j-t)!},\\
\Sigma^{''}_{\boldsymbol{\gamma}}= &
\displaystyle \sum_{t=\max v_i+\frac{1}{2}}^{t=\min p_j}\frac{(-1)^{t}t!}{\Pi_{i}(t-v_i\!-\!\frac{1}{2})!\Pi_{j}(p_j-t)!},
\end{align}
so that
\begin{equation}
\Sigma_{\boldsymbol{\gamma}}= 
\boldsymbol{-}\Sigma^{'}_{\boldsymbol{\gamma }}
\:\boldsymbol{+}\,\displaystyle
\begin{array}{c}
\big [2\sum_{i=1}^{3} j_iJ_i + \frac{1}{2}(\sum_{j=1}^{3}p_j) +\frac{3}{2}\big]\end{array}\Sigma^{''}_{\boldsymbol{\gamma}}.
\end{equation}
In the  first sum $\Sigma^{'}_{\boldsymbol{\gamma }}$ we carry out simultaneously the variable changes $t\rightarrow kx$ and $v_i \rightarrow kv_i, p_j \rightarrow kp_j$
and use eqs. (\ref{Approx1}-\ref{Approx2}). The result is
\begin{equation}\label{eq:FrontalFactGamma'}
\Sigma^{'}_{\boldsymbol{\gamma }}= \frac{1}{(2\pi)^{3}}
\sum_{x=\max v_i+\frac{1}{2}}^{x=\min p_j}G^{'}_{\gamma}(x),
\end{equation}
where
\begin{align}
G^{'}_{\gamma}(x)&= 
e^{\frac{1}{2}\ln \frac{x^{3}}{\Pi_j(p_j-x)}+k\{\imath \pi x+ x\ln(x)-[\sum_i(x-v_i)\ln(x-v_i)+\sum_j(p_j-x)\ln(p_j-x)]\}}
\nonumber \\
&=e^{F_{\gamma'}(x)+k\mathfrak{f}(x)}, \\
F_{\gamma'}(x)&= e^{\frac{1}{2}\ln \frac{x^{3}}{\Pi_j(p_j-x)}},
\end{align}
so that
\begin{equation}
G^{'}_{\gamma}(x)=\frac{\sqrt{x^{3}}}{\sqrt{\Pi_j(p_j-x)}}e^{k\mathfrak{f}(x)}\label{FormulaGamma'}.
\end{equation}
The second sum $\Sigma_{\boldsymbol{\gamma ''}}$ is handled in the same way as the first with slight differences
\begin{equation}\label{eq:FrontalFactGamma''}
\Sigma^{''}_{\boldsymbol{\gamma }}= \frac{1}{(2\pi)^{3}}
\sum_{x=\max v_i+\frac{1}{2}}^{x=\min p_j}G^{''}_{\gamma}(x),
\end{equation}
with 
\begin{align}
&G^{''}_{\gamma}(x)=\nonumber \\
&e^{\frac{1}{2}\ln \frac{1}{k^{2}}\frac{x}{\Pi_j(p_j-x)}+k\{\imath \pi x+ x\ln(x)-[\sum_i(x-v_i)\ln(x-v_i)+\sum_j(p_j-x)\ln(p_j-x)]\}}\nonumber \\
&=\frac{1}{k}\frac{\sqrt{x}}{\sqrt{\Pi_j(p_j-x)}}e^{k\mathfrak{f}(x)}\label{FormulaGamma''}.
\end{align}
		\subsection{Contribution to the saddle points (parity $\gamma$)}\label{ContribSaddleGamma}
From eq. (\ref{FormulaGamma'}) for $\sum^{'}_\gamma$ and the value of $-f''(x_{\pm})$ found by Gurau, see its equation (33),
\begin{equation}
-f''(x_{\pm})= \frac{\mp \imath \sqrt{\bigtriangleup}}{x_{\pm}\Pi_j (p_j-x_{\pm})},
\end{equation}
we derive 
\begin{align} 
&\sqrt{\frac{2\pi }{k(-f''(x_{\pm}))}}\times \frac{\sqrt{x_\pm^{3}}}{\sqrt{\Pi_j(p_j-x_\pm)}} \nonumber \\
&=\sqrt{\frac{2\pi }{k}}
\sqrt{\frac{x_{\pm}}{\mp \imath \sqrt{\bigtriangleup}}}\sqrt{\Pi_j (p_j-x_{\pm})}
\times \frac{\sqrt{x_\pm^{3}}}{\sqrt{\Pi_j(p_j-x_\pm)}}\nonumber \\
&=\sqrt{\frac{2\pi }{k(\mp \imath \sqrt{\bigtriangleup})}}\;x_\pm^{2}. 
\end{align}
The $k$-dependent header factor comes from eq. (\ref{eq:FrontalFactGamma'1}) with $  (2\pi)^{2}\frac{1}{k^{2}}$, from eq. (\ref {eq:FrontalFactGamma'} ) with $ \frac{1}{(2\pi)^{3}}$ and from the equation just above with $\sqrt{\frac{2\pi }{k}}$ what yields finally $\frac{1}{k}\frac{1}{\sqrt{2\pi k^{3}}}$. 
As $\frac{1}{k}\rightarrow dx$ the remaining contributions for $\sum^{'}_\gamma$ are
\begin{equation}
\frac{x_\pm^{2}}{\sqrt{2\pi k^{3}(\mp \imath \sqrt{\bigtriangleup})}}e^{k[\mathfrak {h}(j,J)+\mathfrak{f}(x_\pm)]}.
\end{equation}
For $\sum^{''}_\gamma$ we have the following contribution
\begin{equation}
\frac{1}{k}\frac{x_\pm}{\sqrt{2\pi k^{3}(\mp \imath \sqrt{\bigtriangleup})}}e^{k[\mathfrak {h}(j,J)+\mathfrak{f}(x_\pm)]},
\end{equation}
which must be weighted by its multiplicative coefficient\\
$\big [2\sum_{\iota=1}^{3} j_\iota J_\iota+ \frac{1}{2}(\sum_{j=1}^{3}p_j) +\frac{3}{2}\big]
 \approx k^{2} \sum_{\iota=1}^{3}2 j_\iota J_\iota $. Then $\sum^{'}_\gamma$ is no longer \\ 
dominant that is the only one contribution 
comes from  $\sum^{''}_\gamma$ has the form
\begin{align}
\big[{\sum_{\iota=1}^{3}2 j_\iota J_\iota }\big]
\frac{k^{2}}{k}\frac{x_\pm}{\sqrt{2\pi k^{3}(\mp \imath \sqrt{\bigtriangleup})}}e^{k[\mathfrak {h}(j,J)+\mathfrak{f}(x_\pm)]}
& \nonumber \\
=\big[{\sum_{\iota=1}^{3}2 j_\iota J_\iota }\big]\frac{\sqrt{\pm \imath}\, x_\pm}{\sqrt{48\pi kV}}e^{k[\mathfrak {h}(j,J)+\mathfrak{f}(x_\pm)]}².
\end{align}
As a function of the parameter $A$ defined in eq. (21) in  \cite {Gurau1} for the saddle points equation  we obtain 
\begin{equation}\label{eq:SaddleGammaContrib}
{\tt Sdl}_{\gamma}(x_\pm)=A\frac{\, x_\pm}{\sqrt{48\pi kV}}
e^{\pm \imath  \frac{\pi}{4}+k[\mathfrak {h}(j,J)+\mathfrak{f}(x_\pm)]}.
\end{equation}
From eq. (44) in \cite{Gurau1} the real part of
\begin{equation}
 e^{ 
k[j_1(\mathfrak{h}_{j_1}+\mathfrak{f}_{j_1})+j_2(\mathfrak{h}_{j_2}+\mathfrak{f}_{j_2})
+j_3(\mathfrak{h}_{j_3}+\mathfrak{f}_{j_3})
+J_1(\mathfrak{h}_{J_1}+\mathfrak{f}_{J_1})+J_2(\mathfrak{h}_{J_2}+\mathfrak{f}_{J_2})
+J_3(\mathfrak{h}_{J_3}+\mathfrak{f}_{J_3})]}\nonumber
\end{equation}
is zero, 
then its remains only the imaginary part  (the $h_{j_i,J_i}$ being real) \\
$e^{\pm \imath k[j_1\theta_{j_1}+j_2\theta_{j_2}+j_3\theta_{j_3}+J_1\theta_{J_1}+J_2\theta_{J_2}
+J_3\theta_{J_3}]}$ whence
\begin{align}\label{eq:SaddleGammaCalcul}
{\tt Sdl}_{\gamma}(x_\pm)=&A\frac{\, x_\pm}{\sqrt{48\pi kV}}e^{\pm \imath  \frac{\pi}{4}}\nonumber \\
&\times e^{\pm \imath k[j_1\theta_{j_1}+j_2\theta_{j_2}+j_3\theta_{j_3}+J_1\theta_{J_1}+J_2\theta_{J_2}+J_3\theta_{J_3}]}.
\end{align}
Let us define  a new angle $\varphi_{k_{\gamma}}^{\theta}$ as 
\begin{equation}\label{eq:PHIGammaDef}
\varphi_{k_{\gamma}}^{\theta}=\sum_{\iota =1}^{\iota=3}k(j_\iota \theta_{j_\iota}+J_i\theta_{J_\iota}).
\end{equation}
Then  the final expression of ${\tt Sdl}_{\gamma}(x_\pm)$ writes
\begin{align}
{\tt Sdl}_{\gamma}(x_\pm)=&A\frac{\, x_\pm}{\sqrt{48\pi kV}}
e^{\pm \imath  \{ \frac{\pi}{4}+\varphi_{k_{\gamma}}^{\theta}\}}.
\end{align}
Finally
\begin{align}\label{SumContribGamma+-}
&{\tt Sdl}_{\gamma}(x_{+})+ {\tt Sdl}_{\gamma}(x_{-})=\frac{1}{\sqrt{48\pi kV}}\nonumber \\
&\times
\Big[(Ax_{+}) e^{\imath \big\{   \frac{\pi}{4}+\varphi_{k_{\gamma}}^{\theta}\big\}}
+(Ax_{-}) e^{- \imath \big\{   \frac{\pi}{4}+\varphi_{k_{\gamma}}^{\theta}\big\}\Big]}.
\end{align}\\
Since $(A\,x_{\pm})=\frac{1}{2}[B\pm \imath \sqrt{\bigtriangleup}]$ the sum in (\ref{SumContribGamma+-}) writes as\\
\begin{align}
&{\tt Sdl}_{\gamma}(x_{+})+ {\tt Sdl}_{\gamma}(x_{-})=\frac{1}{\sqrt{48\pi kV}}\nonumber \\
&\times\Big\{Bcos\big( \frac{\pi}{4}+\varphi_{k_{\gamma}}^{\theta}\big)
-24V sin\big( \frac{\pi}{4}+\varphi_{k_{\gamma}}^{\theta}\big)\Big\}.
\end{align}
		\section {{\tt PARITY} $\boldsymbol{\beta}$ and $k$ odd}\label{SectBeta}
\renewcommand{\theequation}{\ref {SectBeta}.\arabic{equation}}
\setcounter{equation}{0}
A useful  correlation table for parity $\boldsymbol{\beta}$ is shown below. \\[0.5em]
\hspace*{1em}
\begin{tabular}{|c|c|c|c|c|}\hline
\multicolumn{5}{|c|}{spin $\mathfrak{j} ^{\ast}_{\beta}$ vertex common  to both triangles $\overline{v},\overline{v}' $ }\\ \hline
 $\overline{v},\overline{v}' \, \frac{1}{2}$ \!integer  & $\mathfrak{j} ^{\ast}_{\beta}$  & $p$  integer
&  $v,v'$  integer & $\overline{p}, \overline{p}'  \, \frac{1}{2}$ \!integer\\ \hline
 $v_4,v_1$ & $j_1$ & $p_1$ & $v_2,v_3$ & $p_2,p_3$\\ \hline
$v_2,v_1$ & $j_2$  & $p_2$ & $v_3,v_4$ & $p_3,p_1$\\ \hline
$v_3,v_1$ & $j_3$  & $p_3$ &$v_4,v_2$ & $p_1,p_2$\\ \hline
$v_2,v_3$ & $J_1$ & $p_1$ & $v_4,v_1$ & $p_2,p_3$\\ \hline
$v_3,v_4$ & $J_2$  & $p_2$ & $v_2,v_1$ & $p_3,p_1$\\ \hline
$v_4,v_2$ & $J_3$  & $p_3$ & $v_3,v_1$ & $p_1,p_2$\\ \hline
\end{tabular}\\[0.8em]
Let us define   $\mathfrak{J} ^{\ast}_{\beta}$ as the column companion of the spin $\mathfrak{j} ^{\ast}_{\beta}$ and represent it by an empty oval.
\noindent An appropriate diagrammatic representation looks as follows\\
\hspace*{0.5em}
\setlength{\unitlength}{0.48cm}
\begin{picture}(22,3)
\linethickness{0.03pt}
\put(10.85,2){\makebox(0,0){\scriptsize \bf $\mathfrak{j} ^{\ast}_{\beta}$ represented by \normalsize$ \bullet$\scriptsize, 
$\mathfrak{J} ^{\ast}_{\beta}$ by an oval,  only  both half-integer  $\overline{v}, \overline{v}'$ are drawn}}
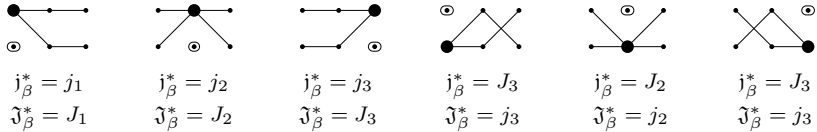
\put(0,0){\circle*{.135}}
 \put(1,1){\circle*{.135}} \put(2,1){\circle*{.135}}
\put(0,1){\circle*{.35}} \put(1,0){\circle*{.135}} \put(2,0){\circle*{.135}}
\put(0,0){\oval(0.35,0.25)}
	\put(0,1){\line(1,-1){1}} \put(1,0){\line(1,0){1}}
	\put(0,1){\line(1,0){1}}\put(1,1){\line(1,0){1}}
\put(1,-1.1){\makebox(0,0){\scriptsize $ {\mathfrak{j} ^{\ast}_{\beta}=j_1}$}}
\put(1,-2){\makebox(0,0){\scriptsize $ {\mathfrak{J} ^{\ast}_{\beta}= J_1}$}}
\put(4,1){\circle*{.135}} \put(5,0){\circle*{.135}} \put(6,1){\circle*{.135}}
\put(4,0){\circle*{.135}} \put(5,1){\circle*{.35}} \put(6,0){\circle*{.135}}
\put(5,0){\oval(0.3,0.25)}
	\put(4,0){\line(1,1){1}} \put(5,1){\line(1,-1){1}}
	\put(4,1){\line(1,0){1}} \put(5,1){\line(1,0){1}}
\put(5,-1.1){\makebox(0,0){\scriptsize $ {\mathfrak{j} ^{\ast}_{\beta}= j_2}$}}
\put(5,-2){\makebox(0,0){\scriptsize $ {\mathfrak{J} ^{\ast}_{\beta}= J_2}$}}
\put(8,1){\circle*{.135}} \put(9,1){\circle*{.135}} \put(10,0){\circle*{.135}}
\put(8,0){\circle*{.135}} \put(9,0){\circle*{.135}} \put(10,1){\circle*{.35}}
\put(10,0){\oval(0.35,0.25)}
	\put(8,0){\line(1,0){1}} \put(9,0){\line(1,1){1}}
	\put(8,1){\line(1,0){1}} \put(9,1){\line(1,0){1}}
\put(9,-1.1){\makebox(0,0){\scriptsize $ {\mathfrak{j} ^{\ast}_{\beta}= j_3}$}}
\put(9,-2){\makebox(0,0){\scriptsize $ {\mathfrak{J} ^{\ast}_{\beta}= J_3}$}}
\put(12,0){\circle*{.35}} \put(13,0){\circle*{.135}} \put(14,0){\circle*{.135}}
\put(12,1){\circle*{.135}} \put(13,1){\circle*{.135}} \put(14,1){\circle*{.135}}
\put(12,1){\oval(0.35,0.25)}
	\put(12,0){\line(1,1){1}} \put(13,1){\line(1,-1){1}}
	\put(12,0){\line(1,0){1}} \put(13,0){\line(1,1){1}}
\put(13,-1.1){\makebox(0,0){\scriptsize $ {\mathfrak{j} ^{\ast}_{\beta}= J_3}$}}
\put(13,-2){\makebox(0,0){\scriptsize $ {\mathfrak{J} ^{\ast}_{\beta}= j_3}$}}
\put(16,0){\circle*{.135}} \put(17,0){\circle*{.35}} \put(18,0){\circle*{.135}}
\put(16,1){\circle*{.135}} \put(17,1){\circle*{.135}} \put(18,1){\circle*{.135}}
\put(17,1){\oval(0.35,0.25)}
	\put(16,0){\line(1,0){1}} \put(17,0){\line(1,1){1}}
	\put(16,1){\line(1,-1){1}} \put(17,0){\line(1,0){1}}
\put(17.1,-1.1){\makebox(0,0){\scriptsize $ {\mathfrak{j} ^{\ast}_{\beta}= J_2}$}}
\put(17.1,-2){\makebox(0,0){\scriptsize $ {\mathfrak{J} ^{\ast}_{\beta}= j_2}$}}
\put(20,0){\circle*{.135}} \put(21,0){\circle*{.135}} \put(22,0){\circle*{.35}}
\put(20,1){\circle*{.135}} \put(21,1){\circle*{.135}} \put(22,1){\circle*{.135}}
\put(22,1){\oval(0.35,0.25)}
	\put(20,1){\line(1,-1){1}} \put(21,0){\line(1,0){1}}
	\put(20,0){\line(1,1){1}} \put(21,1){\line(1,-1){1}}
\put(21.1,-1.1){\makebox(0,0){\scriptsize $ {\mathfrak{j} ^{\ast}_{\beta}= J_3}$}}
\put(21.1,-2){\makebox(0,0){\scriptsize $ {\mathfrak{J} ^{\ast}_{\beta}= j_3}$}}
\end{picture}
\vspace{1.9em}

			\subsection{Change in the prefactor ${\tt R}^{S}_{\beta}$}

Multiplying the spins of the prefactor ${\tt R}^{S}_{\beta}$ by a scaling odd factor $k$ yields
\begin{align}\label{eq:PreFactBeta}
{\tt R}^{S}_{\beta}(k)=(k(p-v))!(k(p-v'))!k((p-\overline{v})-\frac{1}{2})!(k(p-\overline{v}')-\frac{1}{2})! \nonumber\\
\times \,(k(\overline{p}-\overline{v}))!(k(\overline{p}-\overline{v}'))!(k(\overline{p}-v)-\frac{1}{2})!(k(\overline{p}-v')-\frac{1}{2})!\nonumber\\
\times \,(k(\overline{p}'-\overline{v}))!(k(\overline{p}'-\overline{v}'))! (k(\overline{p}'-v)-\frac{1}{2})!(k(\overline{p}'-v')-\frac{1}{2})!\nonumber\\ 
\times \, \big(kv! kv'!\big)^{-1}\big(k\overline{v}+\frac{1}{2})!(k\overline{v}'+\frac{1}{2})!\big)^{-1}.
\end{align}
From eqs. (\ref{Approx2_1}-\ref{Approx3_1}) the numerator of ${\tt R}^{S}_{\beta}(k)$ becomes
\begin{align}
{\tt NumPre\beta}
&=(2\pi)^{6}
e^{[k\ln(k)-k](4\sum_j p_j-3\sum_i v_i)+k[\sum_{j,i}(p_j-v_i)\ln(p_j-v_i)] }\nonumber \\
\times &e^{ \frac{1}{2}[\ln(p-v)+\ln(p-v')+\ln(\overline{p}-\overline{v})+\ln(\overline{p}-\overline{v}')+\ln(\overline{p}'-\overline{v})+\ln(\overline{p}'-\overline{v}')]+3\ln(k)}.
\end{align}
and  the denominator is 
\begin{align}
{\tt DenPre\beta}&=(2\pi)^{2}
e^{[k\ln(k)-k](v+v'+\overline{v}+\overline{v}')}\nonumber \\
&\times e^{k[v\ln(v)+v'\ln(v')+\overline{v}\ln(\overline{v})+\overline{v}'\ln(\overline{v}')]}\nonumber \\
&\times e^{\frac{1}{2}[\ln(kv)+\ln(kv')] 
+\ln(k\overline{v})+\ln(k\overline{v}')}\nonumber \\
&=(2\pi)^{2}
e^{[k\ln(k)-k]\sum_i v_i+k[\sum_i v_i\ln(v_i)]}\nonumber \\
&\times e^{3\ln(k)}e^{\frac{1}{2}\{\ln(v)+\ln(v')\} 
+\ln(\overline{v})+\ln(\overline{v}')}.
\end{align}
Note the lack of frontal factor in $\frac{1}{k^{m}}$ in ${\tt R}^{S}_{\beta}(k)$.\\
Whence
\begin{align}
{\tt R}^{S}_{\beta}(k)& =(2\pi)^{4}
e^{k[\sum_{j,i}(p_j-v_i)\ln(p_j-v_i)-\sum_i v_i\ln(v_i)]}\nonumber \\
& \times
e^{ \frac{1}{2}\ln \big\{
\frac{(p-v)(p-v')(\overline{p}-\overline{v})(\overline{p}-\overline{v}')(\overline{p}'-\overline{v})
(\overline{p}'-\overline{v}')}{(vv')({\overline{v}}^{2}\,{\overline{v}'}^{2})}\big\}}.
\end{align}
Thanks to the correlation table and the definition of the spin $\mathfrak{J} ^{\ast}_{\beta}$ it can be proved that
\begin{equation}
e^{ \frac{1}{2}\ln \big\{
\frac{(p-v)(p-v')(\overline{p}-\overline{v})(\overline{p}-\overline{v}')(\overline{p}'-\overline{v})
(\overline{p}'-\overline{v}')}{(vv')({\overline{v}}^{2}\,{\overline{v}'}^{2})}\big\}}=
\frac{(p-v)(p-v')}{\overline{v}\,\overline{v}'}\,e^{\mathfrak{h}_{\mathfrak{J} ^{\ast}_{\beta}}},
\end{equation}
where the header fraction is dimensionless.
Then 
\begin{align}
{\tt R}^{S}_{\beta}(k) =&(2\pi)^{4}
e^{k[\sum_{j,i}(p_j-v_i)\ln(p_j-v_i)-\sum_i v_i\ln(v_i)]}\nonumber \\
 & \times
\frac{(p-v)(p-v')}{\overline{v}\,\overline{v}'}\,e^{\mathfrak{h}_{\mathfrak{J} ^{\ast}_{\beta}}}.
\end{align}
Finally 
\begin{align}\label{eq:RacRsBeta}
\sqrt{{\tt R}^{S}_{\beta}(k)} =&(2\pi)^{2}
\sqrt{\frac{(p-v)(p-v')}{\overline{v}\,\overline{v}'}}\,e^{\frac{1}{2}\mathfrak{h}_{\mathfrak{J} ^{\ast}_{\beta}}
+k\mathfrak{h}(i,J)}.
\end{align}
			\subsection {Changes in the factors $\Sigma_{\boldsymbol{\beta}}$ and $F_{\beta}(x)$}

From eqs.(\ref{GeneralFormula}) and (\ref{PiMonomS2}) we define two sums $\Sigma^{'}_{\boldsymbol{\beta }}
,\Sigma^{''}_{\boldsymbol{\beta }}$
to be gathered to form a  a $\Sigma_{\boldsymbol{\beta}}$:
\begin{align}
\Sigma^{'}_{\boldsymbol{\beta }}=&
 \sum_{t_{\min}}^{t_{\max}}
\begin{array}{c}
\frac{(-1)^{t}(t+1)!}{(t-v)!(t-v')!
(t-\overline{v}\!-\!\frac{1}{2})!(t-\overline{v}'\!-\!\frac{1}{2})!
(p-t)!(\overline{p}\!+\!\frac{1}{2}-t)!(\overline{p}'\!+\!\frac{1}{2}-t)!}\end{array},\\
\Sigma^{''}_{\boldsymbol{\beta }}=&\sum_{t_{\min}}^{t_{\max}}
\begin{array}{c}
\frac{(-1)^{t}t!}{(t-v)!(t-v')!
(t-\overline{v}\!-\!\frac{1}{2})!(t-\overline{v}'\!-\!\frac{1}{2})!
(p-t)!(\overline{p}\!+\!\frac{1}{2}-t)!(\overline{p}'\!+\!\frac{1}{2}-t)!}\end{array},
\end{align}
so that
\begin{equation}\label{eq:SigmaBetaContrib}
\Sigma_{\boldsymbol{\beta}}= 
\boldsymbol{-}(2 \mathfrak{j} ^{\star}_{\beta}+1)
\Sigma^{'}_{\boldsymbol{\beta }}
\\
\:\boldsymbol{+}\,
[(2 \mathfrak{j} ^{\star}_{\beta}+1)+(\overline{p}+\frac{1}{2})
(\overline{p}' +\frac{1}{2})- vv' ] 
 \,\Sigma^{''}_{\boldsymbol{\beta }},
\end{equation}
with 
\begin{equation}
t_{\min}=\max\!\begin{array}{c}(v,v',\overline{v}\!+\!\frac{1}{2},\overline{v}'\!+\!\frac{1}{2})\end{array}
\; \text{and}\;\, t_{\max}=\min\!\begin{array}{c}(p,\overline{p}\!+\!\frac{1}{2},\overline{p}'\!+\!\frac{1}{2})\end{array}.
\end{equation}
Computation of $\Sigma^{'}_{\boldsymbol{\beta }}$ is very similar to that of the standard sum $\Sigma$ analyzed by Gurau 
whereas $\Sigma^{''}_{\boldsymbol{\beta }}$ looks like
our $\Sigma_{\boldsymbol{\alpha}}$ explicited in sect. {\bf  \ref{SFactorSigmaAlpha}}. \\
With all spins multiplied by the factor $k$,  and changes $t\rightarrow kx, p,v \rightarrow kp,kv$,   $\Sigma^{'}_{\boldsymbol{\beta }}(k)$
rewrites as the sum $\sum_{x=t_{\min}}^{x=t_{\max}}$ over the quotient below
\begin{equation}
\begin{array}{c}
\frac{(-1)^{kx}(kx+1)!}{(k(x-v))!(k(x-v'))!
(k(x-\overline{v})\!-\!\frac{1}{2})!(k(x-\overline{v}')\!-\!\frac{1}{2})!
(k(p-x))!(k(\overline{p}-x)\!+\!\frac{1}{2})!(k(\overline{p}'-x)\!+\!\frac{1}{2})!}\end{array}.
\end{equation}
By using factorial approximations for large $k$ we write $\Sigma^{'}_{\boldsymbol{\beta }}$ in detail as
\begin{equation}
\Sigma^{'}_{\boldsymbol{\beta }}=\frac{1}{(2\pi)^{3}}
\sum_{x=t\min}^{x=t\max}G^{'}_{\beta}(x),
\end{equation}
with
\begin{align}
G^{'}_{\beta}(x)&= e^{[k\ln(k)-k] x  +k\{\imath \pi x+ x\ln(x)\}+\frac{1}{2}\ln(k^{3}x^{3})}\nonumber \\
&\times e^{-\big\{\frac{1}{2}\ln k(x-v)+[k\ln(k)-k] (x-v')+k(x-v')\ln(x-v')+\frac{1}{2}\ln k(x-v')\big\}}\nonumber \\
&\times e^{-\big\{[k\ln(k)-k] (x-v)+k(x-v)\ln(x-v)\big\} }\nonumber \\
&\times e^{-\big\{[k\ln(k)-k](x-\overline{v})+k(x-\overline{v})\ln(x-\overline{v})+[k\ln(k)-k](x-\overline{v}')+k(x-\overline{v}')\ln(x-\overline{v}')\big\} }
\nonumber \\
&\times e^{-\big\{[k\ln(k)-k](p-x)+k(p-x)\ln(p-x)+\frac{1}{2}\ln k(p-x)\big\}}\nonumber \\
&\times e^{-\big\{[k\ln(k)-k](\overline{p}-x)+k(\overline{p}-x)\ln(\overline{p}-x)+\ln k(\overline{p}-x)\big\}}\nonumber \\
&\times e^{-\big\{ [k\ln(k)-k](\overline{p}'-x)+k(\overline{p}'-x)\ln(\overline{p}'-x)+\ln k(\overline{p}'-x)\big\}}.
\end{align}
All rearrangements done it remains
\begin{align}
G^{'}_{\beta}(x)&=\frac{1}{k^{2}}
e^{\frac{1}{2}\ln \frac{x^{3}}{\Pi(x-v_i)\Pi(p_j-x)}
+\frac{1}{2}\ln \frac{(x-\overline{v})(x-\overline{v}')}{(\overline{p}-x)(\overline{p}'-x) }}\nonumber \\
&\times e^{k\{\imath \pi x+ x\ln(x)-\sum  (x-v_i)\ln(x-v_i)-\sum (p_j-x)\ln(p_j-x)\}}.
\end{align}
It results in
\begin{equation}\label{eq;SigmaBeta'}
\Sigma^{'}_{\boldsymbol{\beta }}(k)=
\frac {1}{(2\pi)^{3}}
\sum_{x=t_{\min}}^{x=t_{\max}}\frac{1}{k^{2}}e^{\frac{1}{2}\ln \frac{(x-\overline{v})(x-\overline{v}')}{(\overline{p}-x)(\overline{p}'-x)] }+F_{\beta '}(x)+kf_{\beta '}(x)},
\end{equation}
with
\begin{equation}
F_{\beta '}(x)=\frac{1}{2}\ln \frac{x^{3}}{\Pi(x-v_i)\Pi(p_j-x)}\equiv \mathfrak{F}(x),
\end{equation}
\begin{equation}
f_{\beta '}(x)\equiv \mathfrak{f}(x).
\end{equation}
In the same way we derive
\begin{equation}
\Sigma^{''}_{\boldsymbol{\beta }}(k)=
\frac {1}{(2\pi)^{3}}
\sum_{x=t_{\min}}^{x=t_{\max}}\frac{1}{k^{3}}e^{\frac{1}{2}\ln \frac{(x-\overline{v})(x-\overline{v}')}{(\overline{p}-x)(\overline{p}'-x)] }
+F_{\beta ''}(x)+kf_{\beta ''}(x)},
\end{equation}
with
\begin{equation}
F_{\beta ''}(x)=\frac{1}{2}\ln \frac{x}{\Pi(x-v_i)\Pi(p_j-x)}\equiv  \mathfrak{F}(x)-\ln(x),
\end{equation}
so that
\begin{equation}
e^{F_{\beta ''}(x)}= \frac{1}{x}e^{\mathfrak{F}(x)}\quad \text{and}\quad
f_{\beta ''}(x)\equiv \mathfrak{f}(x).
\end{equation}
			\subsection{Contribution to the saddle points (parity $\beta$)}\label{ContribSaddleBeta}
Identifying $\Sigma^{'}_{\boldsymbol{\beta }}$ of eq. (\ref{eq;SigmaBeta'}) as a Riemann sum with $\frac{1}{k} \rightarrow dx $ leads to
\begin{equation}\label{eq:SaddleContribSigmaBeta'}
\Sigma^{'}_{\boldsymbol{\beta }}(k)=\frac{1}{(2\pi)^{3}k}\int_{t_{\min}}^{t_{\max}}dx \,
e^{\frac{1}{2}\ln \frac{(x-\overline{v})(x-\overline{v}')}{(\overline{p}-x)(\overline{p}'-x)] }+\mathfrak{F}(x)+k\mathfrak{f}(x)}.
\end{equation}
Equation (\ref{eq:SigmaBetaContrib}), with $\mathfrak{j} ^{\star}_{\beta}\rightarrow k\mathfrak{j} ^{\star}_{\beta}$,  gives the following dominant contribution
\begin{equation}
-(2 k\mathfrak{j} ^{\star}_{\beta}+1)\Sigma^{'}_{\boldsymbol{\beta }}(k)=\boldsymbol{-}\frac{2\mathfrak{j} ^{\star}_{\beta}}{(2\pi)^{3}}\int_{t_{\min}}^{t_{\max}}dx \,
e^{\frac{1}{2}\ln \frac{(x-\overline{v})(x-\overline{v}')}{(\overline{p}-x)(\overline{p}'-x)] }+\mathfrak{F}(x)+k\mathfrak{f}(x)}.
\end{equation}
In the same way 
\begin{equation}\label{eq:SaddleContribSigmaBeta''}
\Sigma^{''}_{\boldsymbol{\beta }}(k)=\frac{1}{(2\pi)^{3}k^{2}}\int_{t_{\min}}^{t_{\max}}dx \,\frac{1}{x}\,
e^{\frac{1}{2}\ln \frac{(x-\overline{v})(x-\overline{v}')}{(\overline{p}-x)(\overline{p}'-x)] }+\mathfrak{F}(x)+k\mathfrak{f}(x)}.
\end{equation}
Its coefficient in (\ref{eq:SigmaBetaContrib}),
$[(2 \mathfrak{j} ^{\star}_{\beta}+1)+(\overline{p}+\frac{1}{2})
(\overline{p}' +\frac{1}{2})- vv' ] $,  
is rescaled into
\begin{equation}
[(2 k\mathfrak{j} ^{\star}_{\beta}+1)+(k\overline{p}+\frac{1}{2})
(k\overline{p}' +\frac{1}{2})- k^{2}vv' ]=k^{2}  [\overline{p}\,\overline{p}'-vv' ] +
k[2 \mathfrak{j} ^{\star}_{\beta}+\frac{1}{2}(\overline{p}+\overline{p}')]+\frac {3}{4}.
\end{equation}
The $\Sigma^{''}_{\boldsymbol{\beta }}(k)$ contribution is also dominant and can be written down as \\
\begin{equation}
\frac{1}{x}
\frac {[\overline{p}\,\overline{p}'-vv' ] }{(2\pi)^{3}}
\int_{t_{\min}}^{t_{\max}}dx \,
e^{\frac{1}{2}\ln \frac{(x-\overline{v})(x-\overline{v}')}{(\overline{p}-x)(\overline{p}'-x)] }+\mathfrak{F}(x)+k\mathfrak{f}(x)}.
\end{equation}
Finally the integral approximation has the form
\begin{equation}
\Sigma_{\boldsymbol{\beta}}=\left(-2 \mathfrak{j} ^{\star}_{\beta}+\frac{ [\overline{p}\,\overline{p}'-vv' ] }{x}\right)
\frac {1 }{(2\pi)^{3}}\int_{x=t_{\min}}^{x=t_{\max}}e^{\frac{1}{2}\ln \frac{(x-\overline{v})(x-\overline{v}')}{(\overline{p}-x)(\overline{p}'-x)] } +\mathfrak{F}(x)+k\mathfrak{f}(x)}.
\end{equation}
Writing $\mathfrak{f}(x)$ as the equation (27) used by Gurau for the saddle points
\begin{equation}
\mathfrak{f}(x_{\pm})={j_1 }\mathfrak{f}_{j_1}(x_{\pm})\!+\! j_2 \mathfrak{f}_{j_2}(x_{\pm})
\!+\! j_3 \mathfrak{f}_{j_3}(x_{\pm})
\!+\! J_1 \mathfrak{f}_{J_1}(x_{\pm})\!+\! J_2 \mathfrak{f}_{J_2}(x_{\pm})
\!+\! J_3 \mathfrak{f}_{J_3}(x_{\pm}),
\end{equation}
using its result, see eqs. (26-27-44) in  \cite{Gurau1},
\begin{equation}
\Re(\mathfrak{f}_{\mathfrak{j} ^{\ast}_{\beta}}(x_{+}))=
\Re(\mathfrak{f}_{\mathfrak{j} ^{\ast}_{\beta}}(x_{-}))=
 - \mathfrak{h}_{\mathfrak{j} ^{\ast}_{\beta}}\quad \text{(not depending on }
x_{+}\; \text{or }x_{-}) ,
\end{equation}
our correlation table for $\mathfrak{j} ^{\ast}_{\beta}$ shows that the following identification holds
\begin{equation}
\Re \left\{\ln \frac{(x_{\pm}-\overline{v})(x_{\pm}-\overline{v}')}{(\overline{p}-x_{\pm})(\overline{p}'-x_{\pm})}\right\}
\equiv  \Re \{\mathfrak{f}_{\mathfrak{j} ^{\ast}_{\beta}}(x_{\pm})\}=
 - \mathfrak{h}_{\mathfrak{j} ^{\ast}_{\beta}}.
\end{equation}
In contrast the imaginary part is 
\begin{equation}
\Im\big(\mathfrak{f}_{\mathfrak{j} ^{\ast}_{\beta}}(x_{\pm})\big)= \pm \theta_{\mathfrak{j} ^{\ast}_{\beta}}.
\end{equation}
It follows that
\begin{equation}
\Sigma_{\boldsymbol{\beta}}(x_{\pm})=e^{- \frac{1}{2}\mathfrak{h}_{\mathfrak{j} ^{\ast}_{\beta}}}
\left(-2 \mathfrak{j} ^{\star}_{\beta}+\frac{ [\overline{p}\,\overline{p}'-vv' ] }{x_{\pm}}\right)
\frac {1 }{(2\pi)^{3}}e^{ \pm \frac{\boldsymbol{\imath}}{2}\theta_{\mathfrak{j} ^{\ast}_{\beta}}
+\mathfrak{F}(x_{\pm})+k\mathfrak{f}(x_{\pm})}.
\end{equation}
After the Laplace's approximation the final header factor is simply $\displaystyle \frac {1}{\sqrt{2\pi k}}$.\\
Ultimately from eq. (\ref{eq:RacRsBeta}) the saddle points contributions for parity $\beta$ have the form
\begin{align}
{\tt Sdl}_{\beta}(x_{\pm})=&
\displaystyle \frac {1}{\sqrt{2\pi k}}\sqrt{\frac{(p-v)(p-v')}{\overline{v} \,\overline{v}'}}\, e^{\frac{1}{2}\mathfrak{h}_{\mathfrak{J}^{\star}_{\beta}}}
\nonumber \\
&\times e^{-  \frac{1}{2}\mathfrak{h}_{\mathfrak{j} ^{\ast}_{\beta}}}\,
\frac{\mathfrak{F}(x_{\pm})}{\sqrt{-f''(x_{\pm})}}
\left(-2 \mathfrak{j} ^{\star}_{\beta}+\frac{ [\overline{p}\,\overline{p}'-vv' ] }{x_{\pm}}\right)\nonumber \\
&\times e^{ \pm \frac{\boldsymbol{\imath}}{2}\theta_{\mathfrak{j} ^{\ast}_{\beta}}+k[\mathfrak{h}(i,J)
+\mathfrak{f}(x_{\pm})]}.
\end{align}
From eqs. (32-33) in \cite{Gurau1} clearly  
\begin{equation}
-\mathfrak{f''}(x_{-})= \frac{+\imath\sqrt{\bigtriangleup}}{x_{-}\Pi((p_j-x_{-})}.
\end{equation}
From eq. (35) in \cite{Gurau1} we obtain
\begin{align}
{\tt Sdl}_{\beta}(x_{\pm})=&
\displaystyle \frac {1}{\sqrt{2\pi k}}\sqrt{\frac{(p-v)(p-v')}{\overline{v} \,\overline{v}'}}\, e^{\frac{1}{2}\mathfrak{h}_{\mathfrak{J}^{\star}_{\beta}}}
\nonumber \\
&\times e^{ 
k[j_1(\mathfrak{h}_{j_1}+\mathfrak{f}_{j_1})+j_2(\mathfrak{h}_{j_2}+\mathfrak{f}_{j_2})
+j_3(\mathfrak{h}_{j_3}+\mathfrak{f}_{j_3})]}\nonumber \\
&\times e^{ 
k[J_1(\mathfrak{h}_{J_1}+\mathfrak{f}_{J_1})+J_2(\mathfrak{h}_{J_2}+\mathfrak{f}_{J_2})
+J_3(\mathfrak{h}_{J_3}+\mathfrak{f}_{J_3})]}\nonumber \\
&\times e^{ \frac{1}{2}[\mathfrak{f}_{j_1}+\mathfrak{f}_{j_2}+\mathfrak{f}_{j_3}
+\mathfrak{f}_{J_1}+\mathfrak{f}_{J_2}+\mathfrak{f}_{J_3}]
}\nonumber \\
&\times e^{-  \frac{1}{2}\mathfrak{h}_{\mathfrak{j} ^{\ast}_{\beta}}}\,
\frac{1}{\sqrt{\mp \imath\sqrt{\bigtriangleup}}}
\left(-2 \mathfrak{j} ^{\star}_{\beta}+\frac{ [\overline{p}\,\overline{p}'-vv' ] }{x_{\pm}}\right)
e^{ \pm \frac{\boldsymbol{\imath}}{2}\theta_{\mathfrak{j} ^{\ast}_{\beta}}}.
\end{align}
From $\bigtriangleup=(24)^{2}V^{2}$ and $\frac{1}{\sqrt{\mp \imath}}=e^{\pm \imath\frac{\pi}{4}}$ we derive
\begin{equation}
\frac{1}{\sqrt{\mp \imath\sqrt{\bigtriangleup}}}=\frac{1}{\sqrt{24V}}\,e^{\pm \imath\frac{\pi}{4}}.
\end{equation}
The result for ${\tt Sdl}_{\beta}(x_{\pm})$ becomes the complex expression 
\begin{align}
&{\tt Sdl}_{\beta}(x_{\pm})=
\displaystyle \frac {1}{\sqrt{48\pi kV}}\sqrt{\frac{(p-v)(p-v')}{\overline{v} \,\overline{v}'}}\, e^{\frac{1}{2}\mathfrak{h}_{\mathfrak{J}^{\star}_{\beta}}}\nonumber \\
&\times e^{ 
k[j_1(\mathfrak{h}_{j_1}\!+\!\mathfrak{f}_{j_1})\!+\!j_2(\mathfrak{h}_{j_2}\!+\!\mathfrak{f}_{j_2})
\!+\!j_3(\mathfrak{h}_{j_3}\!+\!\mathfrak{f}_{j_3})
\!+\!J_1(\mathfrak{h}_{J_1}\!+\!\mathfrak{f}_{J_1})\!+\!J_2(\mathfrak{h}_{J_2}+\mathfrak{f}_{J_2})
\!+\!J_3(\mathfrak{h}_{J_3}\!+\!\mathfrak{f}_{J_3})]}\nonumber \\
&\times e^{ \frac{1}{2}[\mathfrak{f}_{j_1}+\mathfrak{f}_{j_2}+\mathfrak{f}_{j_3}
+\mathfrak{f}_{J_1}+\mathfrak{f}_{J_2}+\mathfrak{f}_{J_3}]
}\nonumber \\
&\times e^{-  \frac{1}{2}\mathfrak{h}_{\mathfrak{j} ^{\ast}_{\beta}}}\,
e^{\pm \imath\frac{\pi}{4}}
\left(-2 \mathfrak{j} ^{\star}_{\beta}+\frac{ [\overline{p}\,\overline{p}'-vv' ] }{x_{\pm}}\right)
e^{ \pm \frac{\boldsymbol{\imath}}{2}\theta_{\mathfrak{j} ^{\ast}_{\beta}}}.
\end{align}
From eq. (44) in \cite{Gurau1} the real part of
\begin{equation}
 e^{ 
k[j_1(\mathfrak{h}_{j_1}+\mathfrak{f}_{j_1})+j_2(\mathfrak{h}_{j_2}+\mathfrak{f}_{j_2})
+j_3(\mathfrak{h}_{j_3}+\mathfrak{f}_{j_3})
+J_1(\mathfrak{h}_{J_1}+\mathfrak{f}_{J_1})+J_2(\mathfrak{h}_{J_2}+\mathfrak{f}_{J_2})
+J_3(\mathfrak{h}_{J_3}+\mathfrak{f}_{J_3})]}\nonumber
\end{equation}
is zero, 
then its remains only the imaginary part  (the $h_{j_i,J_i}$ being real) which is\\
$e^{\pm \imath k[j_1\theta_{j_1}+j_2\theta_{j_2}+j_3\theta_{j_3}+J_1\theta_{J_1}+J_2\theta_{J_2}
+J_3\theta_{J_3}]}$ whence
\begin{align}
&{\tt Sdl}_{\beta}(x_{\pm})=
\displaystyle \frac {1}{\sqrt{48\pi kV}}\sqrt{\frac{(p-v)(p-v')}{\overline{v} \,\overline{v}'}}\, e^{\frac{1}{2}\mathfrak{h}_{\mathfrak{J}^{\star}_{\beta}}}\nonumber \\
&\times e^{\pm \imath k[j_1\theta_{j_1}+j_2\theta_{j_2}+j_3\theta_{j_3}+J_1\theta_{J_1}+J_2\theta_{J_2}+J_3\theta_{J_3}]}\nonumber \\
&\times e^{\pm \imath  \frac{1}{2}[\theta_{j_1}+\theta_{j_2}+\theta_{j_3}+\theta_{J_1}+\theta_{J_2}+\theta_{J_3}]
}\nonumber \\
&\times e^{ \frac{1}{2}\Re [\mathfrak{f}_{j_1}+\mathfrak{f}_{j_2}+\mathfrak{f}_{j_3}
+\mathfrak{f}_{J_1}+\mathfrak{f}_{J_2}+\mathfrak{f}_{J_3}]
}\nonumber \\
&\times e^{-  \frac{1}{2}\mathfrak{h}_{\mathfrak{j} ^{\ast}_{\beta}}}\,
e^{\pm \imath\frac{\pi}{4}}
\left(-2 \mathfrak{j} ^{\star}_{\beta}+\frac{ [\overline{p}\,\overline{p}'-vv' ] }{x_{\pm}}\right)
e^{ \pm \frac{\boldsymbol{\imath}}{2}\theta_{\mathfrak{j} ^{\ast}_{\beta}}}.
\end{align}
That is 
\begin{align}
&{\tt Sdl}_{\beta}(x_{\pm})=
\displaystyle \frac {1}{\sqrt{48\pi kV}}
\sqrt{\frac{(p-v)(p-v')}{\overline{v} \,\overline{v}'}}\, 
\nonumber \\
&\times e^{- \frac{1}{2} [\mathfrak{h}_{j_1}+\mathfrak{h}_{j_2}+\mathfrak{h}_{j_3}
+\mathfrak{h}_{J_1}+\mathfrak{h}_{J_2}+\mathfrak{h}_{J_3}
+\mathfrak{h}_{\mathfrak{j} ^{\ast}_{\beta}}-\mathfrak{h}_{\mathfrak{J}^{\star}_{\beta}}]
}\nonumber \\
&\times 
\left([v+v'-\overline{p}-\overline{p}']+\frac{ [\overline{p}\,\overline{p}'-vv' ] }{x_{\pm}}\right)
e^{\pm \imath (\frac{\pi}{4}+\varphi_{k_{\beta}}^{\theta})},
\end{align}
where a new angle $\varphi_{k_{\beta}}^{\theta}$ depending on $\theta_{\mathfrak{j} ^{\ast}_{\beta}}$ is defined {\it via}
\begin{equation}\label{eq:VarPhiBeta}
\varphi_{k_{\beta}}^{\theta}=\sum_{\iota=1}^{\iota=3}[(kj_\iota+\frac{1}{2})\theta_{j_\iota}+(kJ_\iota+\frac{1}{2})\theta_{J_\iota}]
+\frac{1}{2}\theta_{\mathfrak{j} ^{\ast}_{\beta}}.
\end{equation}
We have
\begin{equation}
\frac {1}{x_{+}}= \frac{(B-\imath \sqrt{\bigtriangleup)}}{2C}
\quad \text{and}\quad
\frac {1}{x_{-}}= \frac{(B+\imath \sqrt{\bigtriangleup)}}{2C},
\end{equation}
whence adding the contribution $+$ and $-$, taken into account  $C=\Pi(v_i)$, \\$\sqrt{\bigtriangleup}=24V$, we obtain
\begin{equation}\scriptstyle
\frac{1}{\Pi(v_i)}\Big\{\big\{2\Pi(v_i)[v+v'-\overline{p}-\overline{p}']+B [\overline{p}\,\overline{p}'-vv' ] \big\}\,cos(\frac{\pi}{4}+\varphi_{k_{\beta}}^{\theta})+24V [\overline{p}\,\overline{p}'-vv' ] \, sin(\frac{\pi}{4}+\varphi_{k_{\beta}}^{\theta})\Big\}.
\nonumber \end{equation}
A preliminary result is then
\begin{align}
&\left\{ \begin{array}{ccc} kj_1 &kj_2 & kj_{3} \\ kJ_1 & kJ_2 & kJ_3 \end{array} \right\}_{\beta}^{S}\approx \,
\displaystyle \frac {(-1)^{(v+v'-p)}}{\sqrt{48\pi kV}\,\Pi(v_i)}
\sqrt{\frac{(p-v)(p-v')}{\overline{v} \,\overline{v}'}}
\nonumber \\
&\times e^{- \frac{1}{2} [\mathfrak{h}_{j_1}+\mathfrak{h}_{j_2}+\mathfrak{h}_{j_3}
+\mathfrak{h}_{J_1}+\mathfrak{h}_{J_2}+\mathfrak{h}_{J_3}]
- \frac{1}{2}[\mathfrak{h}_{\mathfrak{j} ^{\ast}_{\beta}}-\mathfrak{h}_{\mathfrak{J}^{\star}_{\beta}}]
}\nonumber \\
&\times \scriptstyle
\Big\{\big\{2\Pi(v_i)[v+v'-\overline{p}-\overline{p}']+B [\overline{p}\,\overline{p}'-vv' ] \big\}\,cos(\frac{\pi}{4}+\varphi_{k_{\beta}}^{\theta})
+24V [\overline{p}\,\overline{p}'-vv' ] \, sin(\frac{\pi}{4}+\varphi_{k_{\beta}}^{\theta})\Big\}.
\end{align}
Factor $ \frac{1}{2}\mathfrak{h}_{\mathfrak{J}^{\star}_{\beta}}$ is compensatory,  {\it i.e.} 
$ \frac{1}{2}\mathfrak{h}_{\mathfrak{J}^{\star}_{\beta}}$ term vanishes whereas 
$ \frac{1}{2}\mathfrak{h}_{\mathfrak{J}^{\star}_{\beta}}$ becomes $\mathfrak{h}_{\mathfrak{J}^{\star}_{\beta}}$.\\
Unfortunately this evidence will be more or less lost in the  general formulas we try to explicit below.\\
Computation of 
$e^{\mathfrak{h}_{\mathfrak{J} ^{\ast}_{\beta}} -  \mathfrak{h}_{\mathfrak{j} ^{\ast}_{\beta}}}$ yields
\begin{align}
e^{\mathfrak{h}_{\mathfrak{J} ^{\ast}_{\beta}} -  \mathfrak{h}_{\mathfrak{j} ^{\ast}_{\beta}}}=
e^{\frac{1}{2}\ln \big\{
\frac{(\overline{p}-\overline{v})(\overline{p}-\overline{v}')
(\overline{p}'-\overline{v})(\overline{p}'-\overline{v}')(p-\overline{v})(p-\overline{v}')\,\overline{v}\,\overline{v}'}
{(p-v)(p-v')(\overline{p}-v)(\overline{p}-v'(\overline{p}'-v)(\overline{p}'-v')\, v\,v' }\big\}},
\end{align}
whence 
\begin{align}
e^{-\frac{1}{2}[\mathfrak{h}_{\mathfrak{J} ^{\ast}_{\beta}} -  \mathfrak{h}_{\mathfrak{j} ^{\ast}_{\beta}}]}=
\scriptsize
\sqrt[4]{\frac{(p-v)(p-v')(\overline{p}-v)(\overline{p}-v'(\overline{p}'-v)(\overline{p}'-v')\, v\,v' }
{(\overline{p}-\overline{v})(\overline{p}-\overline{v}')
(\overline{p}'-\overline{v})(\overline{p}'-\overline{v}')(p-\overline{v})(p-\overline{v}')\,\overline{v}\,\overline{v}'}}.
\end{align}
Besides from eqs. (8-9-11) in \cite{Gurau1} we can write
\begin{equation}
e^{-\mathfrak{H}(i,J)}=e^{-\frac{1}{2}\sum_{j=1}^{3}(\mathfrak{h}_ {j_j}+\mathfrak{h}_ {J_j})}=
\displaystyle e^{-\frac{1}{4}\ln \Pi_{i,j}\frac{(p_j-v_i)}{v_i^{3}}}
=
 \sqrt[4]{ \Pi_{i,j}[v_i^{3}/ (p_j-v_i)]}
\end{equation}
We conclude that
\begin{align}
&\left\{ \begin{array}{ccc} kj_1 &kj_2 & kj_{3} \\ kJ_1 & kJ_2 & kJ_3 \end{array} \right\}_{\beta}^{S}\approx \,
\displaystyle \frac {(-1)^{(v+v'-p)}}{\sqrt{48\pi kV}\,\Pi(v_i)}
\sqrt{\frac{(p-v)(p-v')}{\overline{v}\,\overline{v}'}}
\nonumber \\
&\times \scriptsize \sqrt[4]{ \Pi_{i,j}[v_i^{3}/ (p_j-v_i)]}\, \scriptsize
\sqrt[4]{\frac{(p-v)(p-v')(\overline{p}-v)(\overline{p}-v'(\overline{p}'-v)(\overline{p}'-v')\, v\,v' }
{(\overline{p}-\overline{v})(\overline{p}-\overline{v}')
(\overline{p}'-\overline{v})(\overline{p}'-\overline{v}')(p-\overline{v})(p-\overline{v}')\,\overline{v}\,\overline{v}'}}\nonumber \\
&\times \scriptstyle
\Big\{\big\{2\Pi(v_i)[v+v'-\overline{p}-\overline{p}']+B [\overline{p}\,\overline{p}'-vv' ] \big\}\,cos(\frac{\pi}{4}+\varphi_{k_{\beta}}^{\theta})
+24V [\overline{p}\,\overline{p}'-vv' ] \, sin(\frac{\pi}{4}+\varphi_{k_{\beta}}^{\theta})\Big\}\nonumber \\
&\phantom{\left\{ \begin{array}{ccc} kj_1 &kj_2 & kj_{3} \\ kJ_1 & kJ_2 & kJ_3 \end{array} \right\}_{\beta}^{S}}=\,
\displaystyle \frac {(-1)^{(v+v'-p)}}{\sqrt{48\pi kV}
\, \sqrt[4]{ \Pi_{i,j}[(p_j-v_i)v_i]}}
\nonumber \\
&\times \scriptstyle\sqrt{\frac{(p-v)(p-v')}{\overline{v}\,\overline{v}'}}\, 
\sqrt[4]{\frac{(p-v)(p-v')(\overline{p}-v)(\overline{p}-v'(\overline{p}'-v)(\overline{p}'-v')\, v\,v' }
{(\overline{p}-\overline{v})(\overline{p}-\overline{v}')
(\overline{p}'-\overline{v})(\overline{p}'-\overline{v}')(p-\overline{v})(p-\overline{v}')\,\overline{v}\,\overline{v}'}}\nonumber \\
&\times\scriptstyle
\Big\{\big\{2\Pi(v_i)[v+v'-\overline{p}-\overline{p}']+B [\overline{p}\,\overline{p}'-vv' ] \big\}\,cos(\frac{\pi}{4}+\varphi_{k_{\beta}}^{\theta})
+24V [\overline{p}\,\overline{p}'-vv' ] \, sin(\frac{\pi}{4}+\varphi_{k_{\beta}}^{\theta})\Big\}.
\end{align}
Note that denominator $\sqrt[4]{ \Pi_{i,j}[(p_j-v_i)v_i]}$ has the dimension of an area like $\sqrt{\Pi v_i}$  
 in the formula (\ref{eq:6jSAlphaApprox}) for approximating the supersymmetric symbol $6j^S_{\alpha}$. The following line 
with $\sqrt{}$ and $\sqrt[4]{}$ is dimensionless.


\section{Results for the supersymmetric limits}\label{Results}
\renewcommand{\theequation}{\ref {Results}.\arabic{equation}}
\setcounter{equation}{0}
\noindent
Prior to properly present results as similar as possible with standard  $6j$ symbols we will use a formula to shift the angular arguments, namely:
\begin{equation}
a\cos(x)+b\sin(x)={\cal{N}}cos(x-\psi),
\end{equation}
where
\begin{equation}\label{shiftformula}
{\cal{N}}=\sqrt{a^2+b^2}\quad \mbox{and}\quad \tan(\psi)=\frac{b}{a}.
\end{equation}
Then for each parity  $\boldsymbol{\alpha}$, $\boldsymbol{\gamma}$, $\boldsymbol{\beta}$, we get
\begin{equation}\label{shiftalpha}
{\cal{N}}_\alpha=\sqrt{B^2+(24V)^2},\quad \psi_\alpha= \arctan\Big(\frac{24V}{B}\Big),
\end{equation}
\begin{equation}\label{shiftgamma}
{\cal{N}}_\gamma={\cal{N}}_\alpha, \quad \psi_\gamma=\boldsymbol{-}\psi_\alpha,
\end{equation}
\begin{align}\label{shiftbeta}
{\cal{N}}_\beta=\sqrt{\scriptstyle{(2\Pi(v_i)[v+v'-\overline{p}-\overline{p}']})^2+(24V [\overline{p}\,\overline{p}'-vv' ] )^2}, \\
\psi_\beta= \arctan\Big(\scriptstyle{\frac{24V [\overline{p}\,\overline{p}'-vv' ] }{2\Pi(v_i)[v+v'-\overline{p}-\overline{p}']}}\Big).
\end{align}
We recall that $V$ is the tetrahedron volume and $B$ which also has the dimension of a volume is given by 
\begin{equation} 
B=\Big(\sum_{\iota=1}^{3}j_\iota J_\iota \sum_{j=1}^{3}p_j \Big)+2(j_1j_2j_3 +j_1J_2J_3 +j_2J_3J_1 +j_3J_1J_2).
\end{equation}
\underline{{\tt Parity} $\boldsymbol{\alpha}$}, :\\
{\sl Under a rescaling of all its arguments by a large $k$ a supersymmetric $6j^{S}_\alpha$ symbol behaves like}
\begin{equation} \label{eq:General6jSalpha}
\left\{ \begin{array}{ccc} kj_1 &kj_2 & kj_{3} \\ kJ_1 & kJ_2 & kJ_3 \end{array} \right\}_{\alpha}^{S}=\frac{1}{\sqrt{48\pi kV}\sqrt{\Pi v_i}}
\Big[ {\cal{N}}_\alpha \cos\left(\frac{\pi}{4}+\varphi_{k}^{\theta}-\psi_\alpha\right)\Big],
\end{equation}
where the angle $\varphi_{k}^{\theta} $ is defined by
\begin{equation}\label{eq:VarPhi}
\varphi_{k}^{\theta}=\sum_{\iota=1}^{\iota=3}(kj_\iota+\frac{1}{2})\theta_{j_\iota}+(kJ_\iota+\frac{1}{2})\theta_{J_\iota}.
\end{equation} 
$\theta_{j_\iota}$, $\theta_{J_\iota}$ are the exterior dihedral angles of the tetrahedron corresponding to the edges $j_\iota$ and $J_\iota$ respectively.\\
\noindent
\underline{{\tt Parity} $\boldsymbol{\gamma}$:} [If $k$ is even, refer to formula (\ref{eq:General6jSalpha})]\\ [0.2em]
{\sl Under a rescaling of all its arguments by a large $k$ (odd) a supersymmetric $6j^{S}_\gamma$  behaves like}
\begin{equation} \label{eq:General6jSgamma}
\left\{ \begin{array}{ccc} kj_1 &kj_2 & kj_{3} \\ kJ_1 & kJ_2 & kJ_3 \end{array} \right\}_{\gamma}^{S}=
\frac{(-1)^{(1+\sum p_j)}}{\sqrt{48\pi kV}}
\Big[{\cal{N}}_\alpha \cos\left(\frac{\pi}{4}+\varphi_{k_{\gamma}}^{\theta}+\psi_\alpha\right)
\Big],
\end{equation}
where 
\begin{equation}
\varphi_{k_{\gamma}}^{\theta}=\sum_{\iota=1}^{\iota=3}k(j_\iota \theta_{j_\iota}+J_i\theta_{J_\iota}).
\end{equation}
\noindent
\underline{{\tt Parity} $\boldsymbol{\beta}$:} [If $k$ is even, refer to formula (\ref{eq:General6jSalpha})]\\
{\sl Under a rescaling of all its arguments by a large $k$ (odd)  a supersymmetric $6j^{S}_\beta$  behaves like}
\begin{align}\label{eq:FinalSymbolBeta}
&\left\{ \begin{array}{ccc} kj_1 &kj_2 & kj_{3} \\ kJ_1 & kJ_2 & kJ_3 \end{array} \right\}_{\beta}^{S}=\,
\displaystyle \frac {(-1)^{(v+v'-p)}}{\sqrt{48\pi kV}
\, \sqrt[4]{ \Pi_{i,j}[(p_j-v_i)v_i]}}\sqrt{\frac{(p-v)(p-v')}{\overline{v}\,\overline{v}'}}
\nonumber \\
&\quad\,\times {\scriptstyle \, 
\sqrt[4]{\frac{(p-v)(p-v')(\overline{p}-v)(\overline{p}-v'(\overline{p}'-v)(\overline{p}'-v')\, v\,v' }
{(\overline{p}-\overline{v})(\overline{p}-\overline{v}')
(\overline{p}'-\overline{v})(\overline{p}'-\overline{v}')(p-\overline{v})(p-\overline{v}')\,\overline{v}\,\overline{v}'}}} 
\,\Big[{\cal{N}}_\beta\,cos\Big(\frac{\pi}{4}+\varphi_{k_{\beta}}^{\theta}-\psi_\beta\Big)\Big],
\end{align}
where a new angle depending  on $\theta_{\mathfrak{j} ^{\ast}_{\beta}}$ is  defined as
\begin{equation}
\varphi_{k_{\beta}}^{\theta}=\sum_{\iota=1}^{\iota=3}[(kj_\iota+\frac{1}{2})\theta_{j_\iota}+(kJ_\iota+\frac{1}{2})\theta_{J_\iota}]
+\frac{1}{2}\theta_{\mathfrak{j} ^{\ast}_{\beta}}.
\end{equation}
This means that only one term among the six $(kj_\iota+\frac{1}{2})\theta_{j_\iota}, (kJ_\iota+\frac{1}{2})\theta_{J_\iota}$ transforms into
$(kj_{\mathfrak{j} ^{\ast}_{\beta}}+1)\theta_{\mathfrak{j} ^{\ast}_{\beta}}$. 
Denominator in $\sqrt[4]{}$ has the dimension of an area  like $\sqrt{\Pi v_i}$.

			\section{Conclusion}\label{Conclus}
Compared to a standard $6j$ symbol   [$SU(2)$] a major difference with the supersymmetric $6j^S_{\pi}$ [$OSP(1|2)$]
lies in a  slower decay as a function  of the scaling  parameter $k$. Indeed it becomes $\sqrt{\frac{1}{k}}$ instead of $\sqrt{\frac{1}{k^{3}}}$.
The standard term in $cos$ contains  always angular arguments depending on the six tetrahedral angles $\theta_{j_\iota},\theta_{J_\iota}$.
However all the angles are shifted from their standard values by an angle different according to each parity $\pi$.  \\
For  parity $ \alpha$ the usual expressions like $(kj_\iota+\frac{1}{2})\theta_{j_\iota}$ are unchanged.  For parity $\gamma$ (and $k$ odd) the terms in  $\frac{1}{2}$ vanish so that it remains $(kj_\iota)\theta_{j_\iota}$ and so on.
For parity $\beta$ (and $k$ odd) the angular dependence takes a  form where a special angle $\theta_{\mathfrak{j} ^{\ast}_{\beta}}$ modifies the standard formula into\\
 $(kj_{\mathfrak{j} ^{\ast}_{\beta}}+1)\theta_{\mathfrak{j} ^{\ast}_{\beta}}
+\sum_\iota {(kj_\iota+\frac{1}{2})\theta_{j_\iota}+(kj_\iota+\frac{1}{2})\theta_{J_\iota}}
_{|_{j_\iota,J_\iota \neq\mathfrak{j} ^{\ast}_{\beta} }}$.\\
If $k$ is even all supersymmetric $6j^S_{\pi}$ have the same asymptotic behaviour, {\it ie} that of parity $\alpha$.
Pertinent interpretations are  clearly within the Quantum Supergravity framework.

\section*{Appendix: Expression of supersymmetric prefactors }
\renewcommand{\theequation}{A.\arabic{equation}}
\setcounter{equation}{0}
\begin{align}
{\tt R}^{S}_{\alpha}=&\frac{\prod_{j=1}^{j=3}\prod_{i=1}^{i=4}(p_j-v_i)!}{\prod_{i=1}^{i=4}(v_i)!},
\label{eq:PrefactAlpha}\\
{\tt R}^{S}_{\gamma}=&\frac{\prod_{j=1}^{j=3}\prod_{i=1}^{i=4}(p_j-v_i-\frac{1}{2})!}{\prod_{i=1}^{i=4}(v_i+\frac{1}{2})!},
\label{eq:PrefactGamma}
\end{align}
\begin{align}
\qquad \;\,{\tt R}^{S}_{\beta}=&(p-v)!(p-v')!(p-\overline{v}-\frac{1}{2})!(p-\overline{v}'-\frac{1}{2})! \nonumber \\
&\times \,(\overline{p}-\overline{v})!(\overline{p}-\overline{v}')!(\overline{p}-v-\frac{1}{2})!(\overline{p}-v'-\frac{1}{2})! \nonumber \\
&\times \,(\overline{p}'-\overline{v})!(\overline{p}'-\overline{v}')! (\overline{p}'-v-\frac{1}{2})!(\overline{p}'-v'-\frac{1}{2})! \nonumber\\ 
&\times \, \big(v! v'!\big)^{-1}\big((\overline{v}+\frac{1}{2})!(\overline{v}'+\frac{1}{2})!\big)^{-1}\label{eq:PreFactBetaBis}
\end{align}


\end{document}